\documentclass[%
reprint,
 amsmath,amssymb,
 aip,
 cha,
]{revtex4-2}

\usepackage{graphicx}
\usepackage{dcolumn}
\usepackage{bm}

\usepackage{multirow}
\usepackage{verbatim}
\usepackage{tikz}
\usepackage{pgfplots}
\usepackage{hyperref}
\pgfplotsset{compat=1.18}

\usepackage[normalem]{ulem}
\usepackage{color}

\newcommand{\ba}{\mathbf{a}}
\newcommand{\bP}{\mathbf{P}}
\newcommand{\calE}{\mathcal{E}}
\newcommand{\calP}{\mathcal{P}}
\newcommand{\calW}{\mathcal{W}}
\newcommand{\mutrain}{\mu_\mathrm{train}}
\newcommand{\mutest}{\mu_\mathrm{test}}
\newcommand{\mupred}{\mu_\mathrm{pred}}

\newcommand{\Koop}{\mathcal{K}}
\newcommand{\bK}{\mathbf{K}}
\newcommand{\Lie}{\mathcal{L}}

\newcommand{\Obs}{\mathcal{O}}

\renewcommand{\eqref}[1]{Eq.~\ref{#1}}

\newcommand{\mbf}[1]{\mathbf{#1}}

\newcommand{\reals}{\mathbb{R}} 

\newcommand{\trans}{^\mathrm{T}}

  \newcommand{\bx}{\mbf{x}}
  \newcommand{\by}{\mbf{y}}

  \newcommand{\bL}{\mbf{L}}
  \newcommand{\bI}{\mbf{I}}

  \newcommand{\bff}{\mbf{f}}
  \newcommand{\bfF}{\mbf{F}}

  \renewcommand{\ba}{\mbf{a}}

  \newcommand{\bu}{\mbf{u}}

\newcommand{\bF}{\mathbf{F}}

\newcommand{\bA}{\boldsymbol{A}}

\newlength{\figwidth}
\newlength{\SCwidth}
\def\XXint#1#2#3{{\setbox0=\hbox{$#1{#2#3}{\int}$}
     \vcenter{\hbox{$#2#3$}}\kern-.5\wd0}}

\begin{document}


\title{On the relationship between Koopman operator approximations and neural ordinary differential equations for data-driven time-evolution predictions}
\author{Jake Buzhardt}
\affiliation{Department of Chemical \& Biological Engineering, University of Wisconsin-Madison, Madison, WI 53706, USA}
\author{C. Ricardo Constante-Amores}
\affiliation{Department of Mechanical Science \& Engineering, University of Illinois Urbana-Champaign, Urbana, IL 61801, USA}
\author{Michael D. Graham}%
 \email{mdgraham@wisc.edu}
\affiliation{Department of Chemical \& Biological Engineering, University of Wisconsin-Madison, Madison, WI 53706, USA}

\date{March 14, 2025}

\begin{abstract}
    This work explores the relationship between state space methods and Koopman operator-based methods for predicting the time-evolution of nonlinear dynamical systems. We demonstrate that extended dynamic mode decomposition with dictionary learning (EDMD-DL), when combined with a state space projection, is equivalent to a neural network representation of the nonlinear discrete-time flow map on the state space.  We highlight how this projection step introduces nonlinearity into the evolution equations, enabling significantly improved EDMD-DL predictions. With this projection, EDMD-DL leads to a nonlinear dynamical system on the state space, which can be represented in either discrete or continuous time. This system has a natural structure for neural networks, where the state is first expanded into a high dimensional feature space followed by a linear mapping which represents the discrete-time map or the vector field as a linear combination of these features.  Inspired by these observations, we implement several variations of neural ordinary differential equations (ODEs) and EDMD-DL, developed by combining different aspects of their respective model structures and training procedures. We evaluate these methods using numerical experiments on chaotic dynamics in the Lorenz system and a nine-mode model of turbulent shear flow, showing comparable performance across methods in terms of short-time trajectory prediction, reconstruction of long-time statistics, and prediction of rare events. These results highlight the equivalence of the EDMD-DL implementation with a state space projection to a neural ODE representation of the dynamics. We also show that these methods provide comparable performance to a non-Markovian approach in terms of prediction of extreme events. 
\end{abstract}

\maketitle

\newpage
\begin{quotation}
Accurately predicting the behavior of complex dynamical systems is a critical challenge in many scientific and engineering fields, especially with the increasing availability of large datasets.  
Two promising approaches for developing predictive models from time-series data are neural ODEs, which construct continuous-time models directly on the space of the original data, and approaches based on Koopman operator theory, which lift the data to a high dimensional space where the dynamics can be represented linearly. Recent works have shown that the predictive performance of Koopman-based methods can be significantly improved by projecting the prediction back to the state space on each timestep.  
We show that this step converts the linear dynamics on the space of observables to a nonlinear system on the state space, rendering the method equivalent to a  neural ODE of a particular form. 
This insight helps to form new connections between these seemingly distinct approaches and inspires new methods that combine the strengths of each. 
We validate these findings through numerical experiments and performance comparisons on two chaotic systems arising in fluid dynamics, showing that methods arising from each approach exhibit excellent, and nearly equivalent predictive performance.  
\end{quotation}

\begin{figure*}
    \centering
    \includegraphics[width=0.9\linewidth]{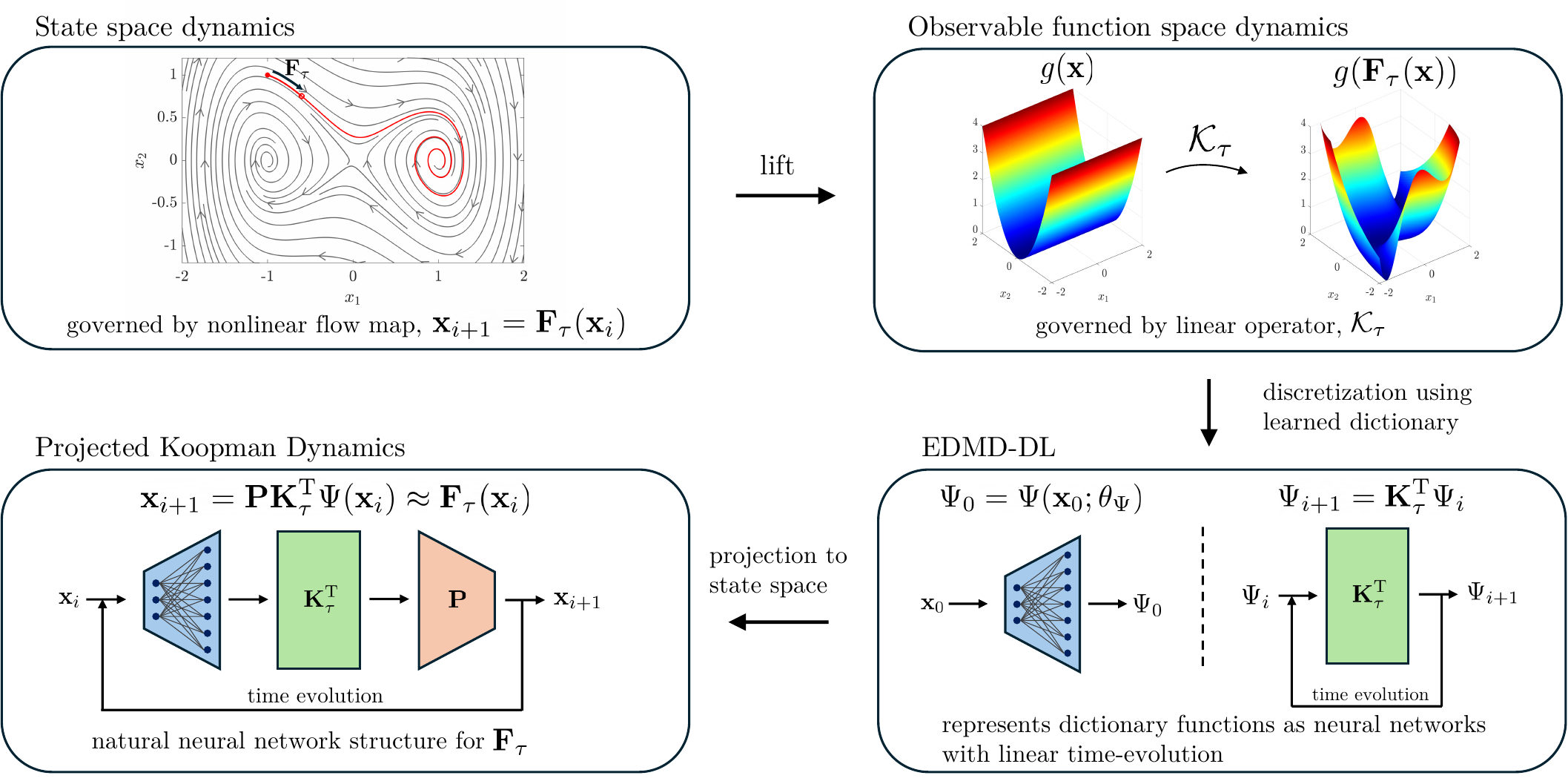}
    \caption{Schematic illustrating how EDMD-DL with a projection to the state space directly leads to a neural network representation of the nonlinear dynamics on the state space.  Here, $\mathbf{x}_i=\mathbf{x}(t_i)$, $\mathbf{x}_{i+1}=\mathbf{x}(t_{i+1})=\mathbf{x}(t_i+\tau)$, and $\mathbf{F}_\tau$ is the flow map for time interval, $\tau$. }
    \label{fig:explanatory}
\end{figure*}

\section{Introduction}
Forecasting the time-evolution of complex dynamical systems from data is of great importance to many scientific and engineering applications, particularly as data availability has increased in recent years.  
This demand has driven the development of neural network-based models for time-series prediction. 
In this work, we establish a connection between two popular approaches to data-driven modeling of dynamical systems:  neural ODEs \cite{Chen2018}, which represent the dynamics directly on the state space, and extended dynamic mode decomposition with dictionary learning (EDMD-DL) \cite{li_extended_2017}, which models the dynamics on the space of observable functions using a finite-dimensional approximation of the Koopman operator. 
Specifically, we show that implementations of EDMD-DL which include a state space projection at each timestep are equivalent to neural ODEs, as this results in a nonlinear neural network-based representation of the dynamical system on the state space.  
This contrasts with the standard Koopman-based approach, which yields a linear representation of the dynamics on the space of observable functions.
The equivalence of the projected Koopman formulation to a neural ODE is illustrated through numerical examples on chaotic systems, where the projected Koopman method achieves predictive performance equivalent to that of a neural ODE while significantly outperforming the standard linear Koopman approach. 
The equivalence between these methods in discrete time is depicted schematically in Fig. \ref{fig:explanatory}. 
Section \ref{sec:continuous_edmd} details this equivalence in continuous time. 

Koopman operator methods aim to provide linear representations of nonlinear dynamical systems by describing time evolution of observable functions \cite{budisic_applied_2012,mezic_analysis_2013,brunton_modern_2022}.
In the Koopman formalism, the dynamics are described by a linear operator, the Koopman operator, which typically acts on an infinite dimensional function space.  
Much of the work in this area has focused on constructing useful finite-dimensional approximations of the Koopman operator directly from data. 
Such approximations hold great promise for many application areas, as they enable the use of modal analysis to identify structure in nonlinear systems \cite{mezic_spectral_2005,mezic_analysis_2013}, they can allow for dimension reduction when dealing with high-dimensional data \cite{rowley_model_2017}, and they can potentially be paired with linear methods for forecasting, estimation, and control \cite{otto_koopman_2021}.  
 
One of the earliest and most common methods to approximate the Koopman operator is dynamic mode decomposition (DMD) \cite{Schmid2010,Rowley2009}, which determines a best-fit linear time-evolution operator given time series data from the system.   
However, the finite dimensional approximation obtained from DMD can only be rigorously connected to the Koopman operator if the observable functions used in the computation lie within an invariant subspace spanned by eigenfunctions of the Koopman operator \cite{mezic_spectral_2005,Rowley2009}.  
In practice, however, this is usually not the case, and 
the lack of an invariant subspace leads to closure issues in approximating the operator. 
This leads to inaccurate approximations of Koopman modes and rapid error buildup when attempting to use the operator approximation for forecasting \cite{brunton_koopman_2016, Wu2021}.  
These issues have led to numerous extensions of the original DMD algorithm \cite{Colbrook2023}, with many of them aiming to improve the function space on which the operator is approximated by a careful selection of observables.  
Extended dynamic mode decomposition (EDMD) \cite{williams_datadriven_2015}, performs the DMD computation on measurement data that is lifted to a higher dimensional space by evaluating a pre-defined set of dictionary functions on the data. 
EDMD with dictionary learning (EDMD-DL)\cite{li_extended_2017} represents the dictionary functions as a neural network and optimizes both the operator matrix and the dictionary to minimize a prediction loss.  
Despite these advances, Koopman-based methods 
with linear time-evolution
still fail to provide accurate forecasts of complex systems (see, for example, 
 Refs. \onlinecite{Fox2023, constante-amores_enhancing_2024} and the results in this paper).
 
Indeed, many Koopman-based methods for forecasting introduce a nonlinearity into the time-evolution formulation to improve performance \cite{li_extended_2017,lusch_deep_2018,otto_linearly_2019,Baddoo2022,Eivazi2021}, with some doing so explicitly and some subtly. 
Of particular interest here is a nonlinearity that has been introduced in several Koopman-based prediction applications, where at each time step in the forward rollout of the observables by the Koopman operator, the observables are projected back to the state space and then ``re-lifted" back to the observable space. 
This procedure was used in the implementation of EDMD-DL for prediction in the work of Li, et al. \cite{li_extended_2017}, and has gained increased attention in recent years.  
J\"unker et al. \cite{junker_data-driven_2022} also pointed out the effectiveness of this method, showing that this strategy  
greatly improves the predictive capability of EDMD as compared to purely linear time evolution of the observables for damped pendulum and Duffing oscillator models.  
Recently, this procedure was further studied in the work of Constante-Amores et al. \cite{constante-amores_enhancing_2024}  
in the context of EDMD-DL, where it was shown that this strategy also yields improved performance for several chaotic systems and a comparison was given in terms of the time between projection steps. 
The method was also considered by van Goor et al. \cite{van_goor_reprojection_2023}, who showed that by viewing the lifting as a mapping to a manifold in a higher dimensional space, the re-lifting procedure can be thought of as a projection back to this manifold, which will not necessarily be invariant under the action of the matrix approximation of the operator.  
Nehma and Tiwari \cite{Nehma2024a} also recently applied this method in their study of learning Koopman representations using Kolmogorov-Arnold networks.

In contrast to the Koopman approach, neural ODEs learn a representation of the dynamics directly on the state space by parameterizing the governing vector field as a neural network. 
The neural ODE method was popularized in recent years by Chen et al. \cite{Chen2018}, who observed that gradients with respect to the network parameters can be computed by either backpropagating through the operations of a numerical ODE solver or by an adjoint sensitivity method.  
These gradients can then be used to optimize the network parameters by stochastic gradient descent.  
While neural ODEs have seen a resurgence in recent years, 
many of the underlying ideas can be found in papers dating back to the 1990s \cite{RicoMartinez1992,RicoMartinez1994,RicoMartinez1993,Chu1991}.

Another popular approach to developing continuous time models directly on the state space is the sparse identification of nonlinear dynamics (SINDy)\cite{Brunton2016}, which models a vector field on the state space as a sparse linear combination of a predefined dictionary of functions of the state (a.k.a. features).  
SINDy is often presented as more interpretable than neural network-based approaches, as the sparsity-promotion in the optimization attempts to eliminate all but the most significant contributors to the dynamics.
However, determining an appropriate set of dictionary functions for a given system, is often challenging and may require system-specific prior knowledge.
Our focus here is on neural ODE models in comparison to EDMD-DL models, as both provide neural network-based modeling strategies to develop models directly from sequences of snapshots. 
We show here that EDMD-DL with a state space projection also represents the vector field as a linear combination of dictionary elements, where the dictionary is represented by a neural network.  In this way, the method can be interpreted as being related to SINDy, without sparsity promotion and on an optimized feature space.

In this work, we elucidate the relationship between EDMD-DL with the projection step and neural ODEs.  
Specifically, we point out that by deviating from the typical, linear time evolution of observables in Koopman formulations, the projection step introduces a nonlinearity into the time evolution, which makes the method equivalent to learning a neural network approximation of the flow map, as illustrated in Fig. \ref{fig:explanatory}.
In this formulation, the flow map approximation has a natural neural network structure that expands out to a high dimensional set of nonlinear features (from EDMD-DL), followed by a linear mapping which selects from those features to represent the flow map. 
Moreover, EDMD-DL parameterizes the dictionary with neural networks, allowing us to draw parallels to neural ODEs.  
We show that converting the discrete time EDMD-DL model to a continuous time formulation directly yields a neural ODE, without the need for a separate training procedure.  
These parallels motivate several model variations developed by combining structures and training methods typically used in neural ODEs and EDMD-DL. 
In comparing the performance of these methods in numerical examples with chaotic dynamics, we further highlight that by including a projection back to the state space on each step of the EDMD-DL time evolution, the resulting model is equivalent to a neural ODE. 
These results clarify an important point regarding the implementation of Koopman operator approximations for time-series prediction --- namely, that a Koopman-based predictor with a state-space projection step provides a representation of the nonlinear discrete-time flow map on the state space and does not preserve the linear time-evolution that is typically associated with Koopman-based methods.  Many works in the literature make a distinction between these methods when, in fact, they are equivalent.    

The remainder of this paper is organized as follows.  
In Sec. \ref{sec:prelim}, we review the dynamical systems formalisms needed for developing models on the state space and based on the Koopman operator.
In Sec. \ref{sec:numerical_methods}, we review numerical methods for training neural network models for dynamical systems and discuss how these models are implemented for prediction tasks.   
In Sec. \ref{sec:results}, we discuss how the parallels between EDMD-DL and neural ODE methods give rise to several variations of these models by combining aspects of each.  We then implement these model variations on two chaotic dynamical systems: the Lorenz system and a nine-mode model for a turbulent shear flow.  With these implementations, we assess the performance of the models in terms of short-time prediction error, reconstruction of long-time statistics, and the prediction of rare events.  

\section{Preliminaries}\label{sec:prelim}
Consider an autonomous dynamical system in continuous time given by an ODE 
\begin{equation}
    \frac{d\bx}{dt}=\bff(\bx).\label{eq:ode}
\end{equation}
where $\bx\in\reals^n$ is the state of the system and denote the associated discrete-time flow map for time interval $\tau$ by $\bF_\tau:\reals^n\to\reals^n$ such that
\begin{equation}\label{eq:flow}
    \bx(t+\tau) = \bF_\tau(\bx(t)). 
\end{equation}
We can also consider time evolution in a space $\Obs$ of observable functions of the state, $g:\reals^n\to\reals$. 
Associated with this system is an operator, 
 $\Koop_\tau:\Obs\to\Obs$ defined by 
\begin{equation}\label{eq:Koopman}
\Koop_\tau g(\bx) = g\circ \bfF_\tau (\bx) = g(\bfF_\tau(\bx))
\end{equation}
which propagates observable functions forward in time along trajectories of the system in \eqref{eq:ode}. This operator $\Koop_\tau$ is known as the Koopman operator, or composition operator \cite{Koopman1931,lasota_chaos_1994}. 
The operator $\Koop_\tau$ is linear, as
\begin{equation}
\begin{split}
    \Koop_\tau(c_1 g_1 + c_2 g_2) (\bx) &= (c_1 g_1 + c_2 g_2)(\bfF_\tau(\bx))\\
    &= c_1g_1(\bfF_\tau(\bx)) + c_2g_2(\bfF_\tau(\bx))\\
    &= c_1\Koop_\tau g_1(\bx) + c_2\Koop_\tau g_2(\bx)
\end{split}
\end{equation}
for $c_1,c_2\in \reals$ and $g_1,g_2\in \Obs$.

The family of operators $\Koop_\tau$ parameterized by $\tau$ has a semigroup structure \cite{lasota_chaos_1994}.  Given sufficiently smooth dynamics, the infinitesimal generator of the Koopman semigroup can be defined and used to express the continuous time form of the dynamics on $\Obs$. 
The infinitesimal generator, $\Lie$ is defined by \cite{lasota_chaos_1994}
\begin{equation} \label{eq:Lie_op}
    \Lie g  = \lim_{t \to 0^+}\frac{g\circ\bF_t-g}{t}
    = \lim_{t \to 0^+}\frac{\Koop_\tau g-g}{t} 
    = \lim_{t \to 0^+}\frac{\Koop_\tau - \mathcal{I}}{t}\,g
    \,. 
\end{equation}
By this definition, the operator $\Lie$ gives the time derivative of $g(\bF_t(\bx))$.  Expanding this by chain rule gives the following explicit form.
\begin{equation}
    \Lie g = \frac{d}{dt}g(\bF_t(\bx)) = \frac{\partial g}{\partial \bx}\cdot\frac{d\,\bF_t(\bx)}{dt} =  \frac{\partial g}{\partial \bx} \cdot \bff(\bx)
    \label{eq:Lie}
\end{equation}
where $\frac{\partial}{\partial \bx}$ denotes the gradient with respect to $\bx$. 
The operator $\Lie$ is commonly referred to as the Koopman generator or the Lie operator \cite{Koopman1931,brunton_modern_2022}, because $\Lie g$ is the Lie derivative of $g$ along the vector field $\bff$~~\cite{Abraham2012}.
Further, by solving \eqref{eq:Lie}, the Koopman operator and Lie operator can be shown to be formally related by an exponential \cite{Cvitanovic2016a}
\begin{equation}\label{eq:Lie_exp}
    \Koop_\tau = e^{\tau \Lie}\,.
\end{equation}
Also, we note that while the Lie operator given in \eqref{eq:Lie} represents a linear partial differential equation, 
this equation is hyperbolic and can be solved by the method of characteristics, wherein each characteristic is determined by solving a nonlinear ODE.  
This is because the Lie operator represents the derivative of the observable in the direction of the flow.
This conceptual point helps to provide some intuition as to how nonlinear dynamics can be represented by a linear operator. 
For further discussion of the conditions under which the generator exists, we direct the reader to the following references: \onlinecite{lasota_chaos_1994, Cvitanovic2016a, Mauroy2020}.

\section{Numerical Methods} \label{sec:numerical_methods}
Here we briefly review
a set of numerical methods for constructing dynamic models from data based on the formalisms introduced in Sec. \ref{sec:prelim}.  
Specifically, we will be interested in learning models from a time series dataset,
 which we will typically organize as snapshot pairs
 \begin{equation} \label{eq:data}
\{(\bx_i,\by_i=\bF_{\tau}(\bx_i))\}_{i=1}^m
\end{equation}
where $\bx_i\equiv \bx(t_i)$, $\tau=t_{i+1}-t_i$ is the sampling interval, and $m$ is the number of data pairs. 
For the continuous time formulations considered below, it is not necessary to have a fixed time interval between datapoints in the dataset, but we will take it to be fixed here for consistency between methods. 
For each of the models considered, we will optimize the appropriate loss over the given dataset.
Therefore, we assume that the dataset is sufficiently large and sampled from a measure on the state space (for example by collecting long trajectories, which effectively samples from a natural measure), so that useful approximations of the dynamics can be inferred from the data.
Under this sampling condition and in the limit of large data, it can be shown that EDMD approximations converge to a Galerkin projection of the Koopman operator \cite{williams_datadriven_2015,klus_numerical_2016,Korda2018}.

While we will be working with a clean, noise-free dataset of full-state observations uniformly spaced in time, we note that other works have considered extensions of the neural ODE framework to more realistic data scenarios.
For example, in cases where data is gathered with irregular sampling times, it is still straightforward to train a neural ODE model \cite{Chen2018}, as the numerical integration can return model predictions at arbitrary points in time. 
Additionally, Ref. \onlinecite{Young2023} showed that these sorts of neural ODE and discrete-time neural network models can be extended to cases where only partial state observations of a chaotic system are available by making use of Takens' embedding theorem and time-delay embeddings.  
Other methods such as latent ODEs \cite{Chen2018} and neural controlled differential equations \cite{Kidger2020} have been developed to extend neural ODE models to realistic scenarios, such as missing data, partial observations, irregular sampling, and incoming data. 
With regards to sensor noise, neural ODEs have been shown to perform well in cases of moderate noise added to the data without altering the training procedure \cite{Chen2018,Kidger2021}. 
In cases of severe sensor noise, other steps could be taken, such as smoothing the data with a filter as a pre-processing step or adding additional regularization to the neural ODE training to discourage overfitting  (see, for example, \onlinecite{Kidger2021,Finlay2020,Kelly2020})

We will additionally be interested in fitting models using neural networks as function approximators, and those which preserve the Markovian structure assumed in the dynamical systems formulation above -- that future time evolution is determined only from the present states.  This requirement excludes several popular deep learning based time series modeling approaches, such as reservoir computing, LSTMs, and transformers.  

Continuous-time and discrete-time models each have their own advantages and disadvantages. In many cases, especially when dealing with physical systems governed by continuous-time equations of motion, a continuous-time model is preferred as it more naturally represents continuous-time physics. Additionally, continuous-time models offer greater versatility in terms of temporal spacing of the training data and predictions, whereas discrete-time models are typically constrained to a single fixed timestep.
It has also been shown that neural ODE methods tend to outperform discrete-time models in cases where the data is widely spaced in time \cite{Linot2022}, which is the case in many real-world data scenarios. 
However, in applications where model runtime is critical and predictions are needed only at a constant sampling frequency, discrete-time models can be advantageous. These models often provide faster inference since they generally require fewer function evaluations per timestep than neural ODEs, which typically use more sophisticated timestepping schemes.  

\subsection{Neural ODEs}\label{sec:neural_odes} 
Developing a continuous-time model
requires an approximation of the ODE vector field $\bff(\bx)$ in \eqref{eq:ode}. 
When $\bff$ is approximated by a neural network $\bff(\bx; \theta_\bff)$, the resulting model is called a neural ODE \cite{Chen2018}. 
The parameters of a neural ODE are optimized to minimize the objective 
\begin{equation}\label{eq:node_objective}
    J(\theta_\bff)= \sum_{i=1}^m \left\|\by_i -  \hat{\by}_i\right\|_2^2 ,
\end{equation} 
where $\hat{\by}_i$ is the prediction of $\by_i$ found by integrating the ODE from $t_i$ to $t_i+\tau$ from initial condition $\bx_i$ 
\begin{equation}\label{eq:node_solve}
    \hat{\by}_i = \bx_i + \int_{t_i}^{t_i+\tau}\bff(\bx(t); \theta_\bff) dt\,.
\end{equation}
In practice, \eqref{eq:node_solve} is evaluated by numerically integrating from a given initial condition.  
In this work, we use a standard Dormand-Prince 4(5) algorithm implemented in the \texttt{torchdiffeq} \cite{torchdiffeq} library using Python and PyTorch \cite{Ansel_PyTorch_2_Faster_2024}.  
The objective in \eqref{eq:node_objective} is typically optimized by stochastic gradient descent, where the gradients with respect to the parameters $\theta_\bff$ are determined by either solving an adjoint equation backward in time from $t_i+\tau$ to $t_i$ to obtain these sensitivities or by using modern automatic differentiation tools to backpropagate these gradients through the operations of a numerical ODE solver \cite{Chen2018}. 
Once the neural ODE model is trained, it can be implemented to predict the time evolution of a system by simply integrating along the learned vector field using a numerical ODE solver from a given initial condition.

\subsection{EDMD with dictionary learning}
Extended dynamic mode decomposition (EDMD) \cite{williams_datadriven_2015} constructs a matrix  approximation of the Koopman operator, $\bK_\tau$
by solving a least squares problem to minimize the objective 
\begin{equation}\label{eq:min_K_edmd_sum}
J(\bK_\tau) = \sum_{i=1}^m\left\| \Psi(\by_i) - \bK_\tau\trans\Psi(\bx_i)\right\|_2^2
\end{equation}
over the dataset of interest for a given vector of dictionary functions, $\Psi$.  
For full details on the EDMD minimization procedure, we refer the reader to Ref. \onlinecite{williams_datadriven_2015}. 

A challenge of the EDMD approach is finding an appropriate dictionary on which to compute the operator $\bK_t$, as the as the accuracy and convergence of the approximation depend heavily on this choice.  
Ideally, the subspace spanned by the dictionary should be Koopman invariant so that the residual can be driven to zero.  
To meet this requirement, it is necessary that the dictionary elements all lie in a space spanned by Koopman eigenfunctions \cite{mezic_spectral_2005, Rowley2009}.
Additionally, we would like a dictionary from which the states can be recovered easily, and preferably linearly, for use in prediction applications.  
However, finding a dictionary that satisfies these properties is nontrivial, as the appropriate dictionary depends heavily on the underlying dynamical system and many common choices of dictionary do not scale well with the state dimension, making them challenging to apply to systems of dimension greater than 3 or 4.  

An approach proposed by Li et al. \cite{li_extended_2017} is to allow the vector of dictionary elements $\Psi$ to be represented by a neural network, parameterized by weights $\theta_\Psi$; that is, 
    $\Psi(\bx) = \Psi(\bx; \theta_\Psi)$. 
This allows the basis to be optimized jointly along with the matrix approximation of the Koopman operator, $\bK_\tau$, so that the objective in \eqref{eq:min_K_edmd_sum} becomes 
\begin{equation}\label{eq:min_K_EDMD_DL}
    J(\bK_\tau, \theta_\Psi) = \sum_{i=1}^m\left\|\Psi(\by_i; \theta_\Psi) - \bK_\tau\trans\Psi(\bx_i; \theta_\Psi) \right\|_2^2
\end{equation}
This approach is known as EDMD with dictionary learning (EDMD-DL).  
The original paper of Li et al \cite{li_extended_2017} proposed an iterative procedure for solving \eqref{eq:min_K_EDMD_DL} which alternates between two steps in which (1) the dictionary is fixed and the Koopman approximation $\bK_\tau$ is optimized through the EDMD and then (2) the matrix $\bK_\tau$ is fixed and the dictionary is optimized through stochastic gradient descent.  
Recently, Constante-Amores et al. \cite{constante-amores_enhancing_2024} showed that these steps can be combined by performing the least-squares calculation for $\bK_\tau$ within the dictionary optimization step and backpropagating
through both the lifting and the least-squares EDMD optimization of $\bK_\tau$. 
 Other works have optimized the elements of $\bK_\tau$ directly along with the dictionary by stochastic gradient descent, avoiding the least squares calculation entirely (e.g., \onlinecite{otto_linearly_2019,lusch_deep_2018,folkestad_koopnet_2022}). 

\subsection{Forecasting with EDMD models}\label{sec:edmd_forecasting}
To forecast the evolution of the states, $\bx$ from an initial condition, $\bx_0$ at $t=0$, we begin
by using $\bK_t$ to evolve the dictionary elements $\Psi(\bx)$ forward in time, as in the objective in \eqref{eq:min_K_edmd_sum}, 
\begin{equation}\label{eq:linprop_psi}
\Psi(\bx(\tau)) = \bK_\tau\trans \Psi(\bx_0) .
\end{equation} 
While this method propagates the features $\Psi$ linearly, it is still necessary to recover the states $\bx$ from the predictions of $\Psi$. 
That is, we need a mapping 
$\calP:\reals^n\to\reals^d$,
which effectively inverts the lifting $\Psi$, projecting the features back to the state space; i.e. $\bx = \calP\Psi(\bx)$.  
One common approach to this is to structure the dictionary so that the states $\bx$ are included explicitly \cite{li_extended_2017,constante-amores_enhancing_2024}; that is, 
\begin{equation}\label{eq:state_in_dict}
    \Psi(\bx) = \begin{bmatrix} \bx \\ \tilde{\Psi}(\bx) \end{bmatrix}  
\end{equation}
where $\tilde{\Psi}$ are dictionary elements which are nonlinear functions of $\bx$.
With this structure, the projection back to the state space takes the simple, linear form $\calP\Psi(\bx) = \bP\Psi(\bx)$ with $\bP = \begin{bmatrix}
    \mathbf{I}_{n\times n} & \mathbf{0}_{n\times(k-n)}
\end{bmatrix}$. 
Alternatively, if the states are not explicitly included in the dictionary, the mapping $\bP$ can be learned as a linear mapping in a similar manner to $\bK_\tau$; that is, by a least squares calculation for standard EDMD \cite{korda_linear_2018} or jointly by gradient descent for EDMD-DL.  
A more general approach is to assume that $\calP$ is a nonlinear mapping and represent it as a decoder neural network, as in many Koopman autoencoder formulations \cite{lusch_deep_2018,otto_linearly_2019}.  Otto and Rowley \cite{otto_linearly_2019} performed comparisons of nonlinear reconstruction versus linear reconstruction of the observables from the features, showing that for some examples performance actually degrades by allowing for nonlinear reconstruction.  
For this reason, in our implementations of EDMD-DL, we will include the state explicitly in the dictionary, as in \eqref{eq:state_in_dict}. 
Including the states $\bx$ as nontrainable dictionary elements also prevents the training procedure from learning the trivial solution, $\Psi(\bx)=\mathbf{0}$, which is a minimizer of \eqref{eq:min_K_EDMD_DL}\cite{li_extended_2017}. 
It is also useful to normalize the error in the state prediction and the dictionary prediction in the loss function\cite{otto_linearly_2019}
\begin{equation}  \label{eq:edmd_loss}
\begin{split}
    J(\bK_\tau; \theta_\Psi) &= \sum_{i=1}^m\Bigg[\frac{\|\by_i - \bP\bK_\tau\trans\Psi(\bx_i;\theta_\Psi) \|_2}{\| \by_i\|_2+\epsilon}\\
    &\qquad \qquad +
    \lambda \, \frac{\|\tilde{\Psi}(\by_i;\theta_\Psi) - \tilde{\bP}\bK_\tau\trans\Psi(\bx_i; \theta_\Psi)\|_2}{\|\tilde{\Psi}(\by_i;\theta_\Psi)\|_2+\epsilon}
    \Bigg]
\end{split}
\end{equation}
where 
$
\tilde{\bP}
=
\begin{bmatrix}
    \mathbf{0}_{(k-n)\times n} & \mathbf{I}_{(k-n)\times (k-n)}
\end{bmatrix}
$ 
is the matrix which extracts only the nonlinear functions $\tilde{\Psi}$ from the dictionary (but not the states $\bx$), $\epsilon$ is a small parameter introduced to avoid divide-by-zero errors, and $\lambda$ is an optional weighting parameter to adjust the relative effect of the two terms. 
The reasoning for this is that the objective in \eqref{eq:min_K_EDMD_DL} can be reduced by simply choosing $\theta_\Psi$ to reduce the magnitude of $\Psi(\bx;\theta_\Psi)$, which may or may not improve the state prediction.  

\subsubsection{EDMD-DL with projection}

The standard time-evolution strategy using an EDMD model, as described above, is to lift the initial state $\bx_0$ to obtain the initial feature vector $\Psi_0 = \Psi(\bx_0)$ and then propagate this lifted state forward linearly using the matrix approximation $\bK_\tau$ as in \eqref{eq:linprop_psi}, 
\begin{equation} \label{eq:linprop_psi_i}
    \Psi_{i+1} = \bK_\tau\trans\Psi_i
\end{equation}
where the subscripts on $\Psi$ refer to values at timesteps evenly spaced by time interval, $\tau$. 
Iterating this for multiple timesteps allows us to obtain a trajectory in the feature space.  
Once this trajectory is obtained, the corresponding trajectory of the states, $\bx$ can be obtained by mapping back to the state space using the operator $\calP$.
A key characteristic of Koopman-based approaches is that, as long as the operator $\calP$ is linear, these methods allow for state trajectories to be predicted from an initial condition using only linear operations (following the initial nonlinear mapping to the feature space).  
This is key, as preserving linearity is necessary in order to apply linear methods for control, estimation, or other downstream tasks \cite{korda_linear_2018}. 

However, several recent works \cite{junker_data-driven_2022,constante-amores_enhancing_2024,constante-amores_data-driven_2024,van_goor_reprojection_2023} have shown that the predictive capabilities of EDMD models can be substantially improved by performing a projection during the forward rollout where after each step with the Koopman model, the prediction is projected from the lifted state $\Psi(\bx_i)$ back to the original state space, 
\begin{equation}\label{eq:nonlinprop_psi_i}
\bx_{i+1} = \bP\bK_\tau\trans \Psi(\bx_i) \,.
\end{equation}
To understand why this subtle change offers such improved performance, we first note that \eqref{eq:nonlinprop_psi_i} can be thought of as a nonlinear discrete-time model on the state space with a particular form, as depicted in Fig. \ref{fig:explanatory}.  
That is, the model in \eqref{eq:nonlinprop_psi_i} 
approximates the nonlinear discrete-time flow map on the state space as 
\begin{equation}
    \bF_\tau(\bx; \theta_\Psi) = \bP\bK_\tau\trans\Psi(\bx;\theta_\Psi). 
\end{equation}
So, in EDMD-DL, a set of features $\Psi$ is learned, which is typically of higher dimension than the original state representation, $\bx$, and whose time evolution is approximately linear.
However, implementing this model for state prediction with a projection on each step defines a nonlinear discrete time map on the state space.  
This map first expands out to a high-dimensional feature vector, followed by two linear mappings: the first, $\bK_\tau\trans$, represents the time evolution and the second, $\bP$, represents a projection back to the original state space.
Since it is common for the output layer of a neural network to be linear (with no activation function following it), this model is essentially a neural network representation of the flow map in a fairly standard form.  This observation is the heart of the present work.  

\begin{figure*}
    \centering
    \includegraphics[width=0.9\linewidth]{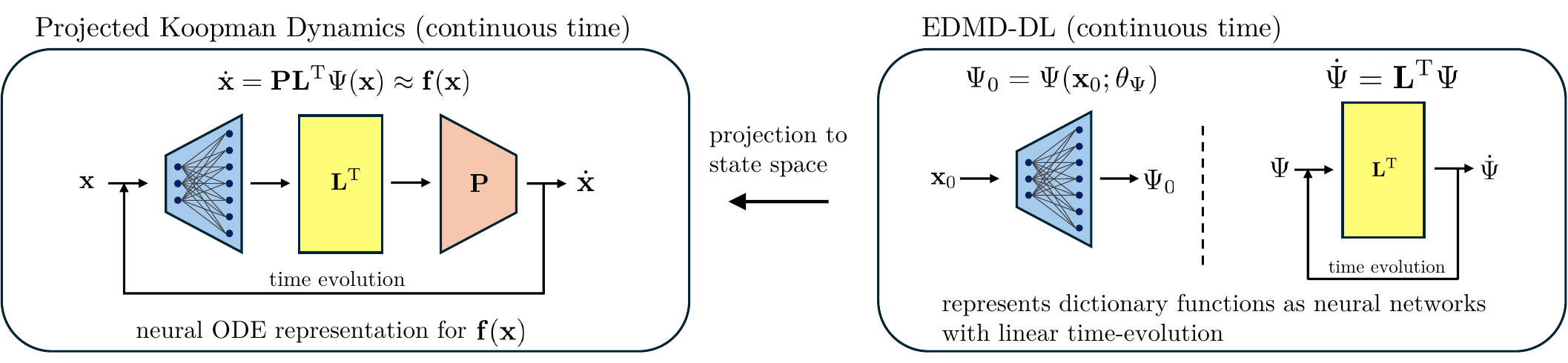}
    \caption{Schematic illustrating how the continuous-time EDMD-DL formulation with projection to the state space leads to a neural ODE representation of the state space dynamics. }
    \label{fig:continuous_EDMD}
\end{figure*}
\subsection{Continuous-time models from EDMD} \label{sec:continuous_edmd}
As noted in Sec. \ref{sec:prelim}, the Koopman formalism also gives rise to a continuous-time representation of the dynamics on the observable space, described by the infinitesimal generator, $\Lie$.  
Based on this, several previous works have shown that continuous-time representations can also be derived from EDMD or similar regression methods \cite{mauroy_linear_2016,Mauroy2020,klus_data-driven_2020,peitz_data-driven_2020, Guo2024}.
Using this continuous-time formulation, an analogous projection step can be performed, which directly yields a neural ODE representation of the state space dynamics, as illustrated in Fig. \ref{fig:continuous_EDMD}. 

In many implementations, it is assumed that time derivative data of the states  $\dot{\bx}$ is available (or can be obtained by finite-differencing).  If this velocity data is available, 
then a matrix representation $\bL$ of the Koopman generator, $\Lie$ can be obtained 
by minimizing the objective 
\begin{equation}
J(\bL) = \sum_{i=1}^m\left\| \dot{\Psi}(\bx_i) - \bL\trans\Psi(\bx_i)\right\|_2^2
\end{equation}
where $\dot{\Psi}(\bx)=\frac{\partial \Psi}{\partial \bx}\cdot\dot{\bx}$. 
This is the approach taken in \cite{klus_data-driven_2020,peitz_data-driven_2020, Guo2024}.  

Alternatively, one can obtain a matrix approximation of the generator from the EDMD approximation of the Koopman operator by applying the limit definition in \eqref{eq:Lie_op} as 
\begin{equation}
    \bL = \frac{\bK_\tau-\bI_{k\times k}}{\tau}
\end{equation}
though this approach will only be accurate for sufficiently small time interval, $\tau$.  
Similarly, the exponential relationship between the Koopman operator and the generator in \eqref{eq:Lie_exp} can be used to obtain a matrix approximation of $\Lie$ using a matrix logarithm as 
\begin{equation} \label{eq:gedmd_log}
    \bL = \frac{1}{\tau}\log(\bK_\tau)\,.
\end{equation}
This is the approach taken by Mauroy and Goncalves \cite{mauroy_linear_2016,Mauroy2020}, where it was also shown that an ODE vector field can be computed directly from an EDMD approximation of the Koopman operator by solving a linear least squares problem for systems linear in parameters.  However, those works were limited to polynomial vector fields and vector fields composed of a linear combination of predefined dictionary functions, respectively. 

Once the matrix representation of the generator is obtained, it can be used to predict the dynamics of the state, $\bx$, as this matrix acts on the dictionary elements to return their time derivative.  
Specifically, an initial state, $\bx_0$, can be lifted to obtain an initial condition for the lifted state, $\Psi_0 = \Psi(\bx_0)$, and then a trajectory in this lifted state can be obtained by integrating the equation
\begin{equation} \label{eq:linprop_L}
    \frac{d\Psi}{dt} = \bL\trans\Psi\,.
\end{equation}
Integrating this yields a trajectory in the lifted state, $\Psi$, from which we can recover the state predictions by mapping back to the state space, $\bx = \calP \Psi$, as in Sec. \ref{sec:edmd_forecasting}. 
Note that Eq. \ref{eq:linprop_L} is a linear ODE in $\Psi$ and is essentially a continuous time version of the linear time evolution using the Koopman operator (without projection) in Eqs. \ref{eq:linprop_psi}, \ref{eq:linprop_psi_i}.

Analogously to the EDMD formulation, where a nonlinear discrete-time evolution equation on the state space was defined using the EDMD operator, $\bK_\tau$, by projecting back to the state space on each timestep (see \eqref{eq:nonlinprop_psi_i}), a nonlinear continuous time evolution equation on the state space can also be defined using the EDMD generator, $\bL$.  This is done by lifting the state $\Psi(\bx)$, computing the time derivative of the feature vector, and then projecting the time derivative back to the state space.  
This can be seen by simply taking the time derivative of the projection back to the state space $\bx = \calP\Psi(\bx)$ as 
\begin{equation}  \label{eq:edmd_ode_general}
    \frac{d\bx}{dt} 
    = \frac{\partial \calP}{\partial \Psi} \frac{d\Psi}{dt} 
    = \frac{\partial \calP}{\partial \Psi} \bL\trans\Psi(\bx)\,.  
\end{equation}
This form is only necessary if the mapping $\calP$ back to the state space is nonlinear (such as in Koopman autoencoder formulations).  
If $\calP$ is taken to be a linear mapping, $\calP\Psi=\bP\Psi$, such as when the state is explicitly included in the dictionary (see Sec. \ref{sec:edmd_forecasting}, \eqref{eq:state_in_dict}), then $\frac{\partial \calP}{\partial \Psi} = \bP$ and \eqref{eq:edmd_ode_general} can be simplified to 
\begin{equation}\label{eq:edmd_ode}
    \frac{d\bx}{dt} = \bP\bL\trans\Psi(\bx)\,.
\end{equation}
In the case of EDMD-DL, where the feature vector $\Psi(\bx)$ is parametrized by a neural network, $\Psi(\bx; \theta_\Psi)$ and optimized jointly with $\bK_\tau$, this implies that the EDMD-DL training procedure directly defines a neural ODE of the form
\begin{equation} \label{eq:edmd_node}
    \frac{d\bx}{dt} = \bP\bL\trans\Psi(\bx; \theta_\Psi)
    \,.
\end{equation}
In particular, the structure of this neural ODE is such that the state is expanded out to a high dimensional set of features, $\Psi(\bx; \theta_\Psi)$ which have been chosen as important during EDMD-DL training, and then the linear mapping $\bP\bL\trans$ selects 
a linear combination of these features to represent the vector field.

In summary, in both discrete time and continuous time, performing a projection back to the state space in the EDMD time evolution leads to a nonlinear dynamical system in the state space with a natural structure for neural networks.  Specifically, the state is first expanded nonlinearly into a high dimensional feature space, followed by a linear mapping which represents the dynamics as a linear combination of these features.  
This observation motivates the development of several methods that combine different aspects of EDMD and neural ODE based models. 
In the following section, we show that adding the projection step to the EDMD-DL time evolution leads to much better predictive performance than the linear time evolution strategy used in standard EDMD-DL. 
Additionally, the models using this projection achieve a level of accuracy in predicting extreme events comparable to the non-Markovian approach of Racca and Magri \cite{Racca2022}. 

\begin{table*}
\setlength{\tabcolsep}{7pt}
\renewcommand{\arraystretch}{0.9}
    \centering
    \caption{Summary of model structures based on neural ODEs and EDMD-DL.}
    \resizebox{\textwidth}{!}{
    \renewcommand{\arraystretch}{1.5} 
    \begin{tabular}{l|c| l}
        \hline
        {\it Model} & {\it Time evolution} & {\it  Training notes} \\
        \hline
        \hline
        Basic neural ODE &  $\dot{\bx} = \bff(\bx; \theta_\bff)$& $\theta_\bff$ optimized as neural ODE\\
        &$\bx(0)=\bx_0$ &   (loss in \eqref{eq:node_objective})\\
        \hline 
        EDMD structured neural ODE \qquad &  $\dot{\bx} = \bA\Psi(\bx; \theta_\Psi)$ & $\theta_\Psi$ and matrix $\bA=\bP\bL\trans$ optimized  \\
        & $\bx(0)=\bx_0$ & as neural ODE  (loss in \eqref{eq:node_objective})\\
        \hline
        EDMD-DL & $\Psi_{i+1} = \bK\trans\Psi_i $  & $\theta_\Psi$ and $\bK$ optimized by EDMD-DL \\
                & $ \bx_i = \bP\Psi_i$ & (loss in \eqref{eq:edmd_loss})\\
                & $\Psi_0 = \Psi(\bx_0; \theta_\Psi)$ \\
        \hline
        EDMD-DL with projection & $\bx_{i+1}=\bP\bK\trans\Psi(\bx_i; \theta_\Psi)$ & $\theta_\Psi$ and $\bK$ optimized by EDMD-DL  \\
        & & (loss in \eqref{eq:edmd_loss})\\
        \hline
        EDMD basis neural ODE & $\dot{\bx} = \bA\Psi(\bx; \theta_\Psi)$ & $\theta_\Psi$ optimized by EDMD-DL (\eqref{eq:edmd_loss})\\
        & $\bx(0)=\bx_0$ & $\bA=\bP\bL\trans$ optimized as neural ODE (\eqref{eq:node_objective})\\
        \hline
        EDMD direct neural ODE & $\dot{\bx} = \bP\bL\trans\Psi(\bx; \theta_\Psi)$ & $\theta_\Psi$, $\bK$ optimized by EDMD-DL (\eqref{eq:edmd_loss}) \\
        & $\bx(0)=\bx_0$ & $\bL=\frac{1}{\tau}\log{\bK}$ \\
        \hline
    \end{tabular}
    }
    \label{tab:knode_models}
\end{table*}
\section{Performance comparison} \label{sec:results}

The discussion in Sec. \ref{sec:numerical_methods} of the parallels between data-driven state-space modeling techniques and the projected Koopman approach to predicting the state evolution motivates a comparison between the numerical methods arising from each.  
In particular, we will compare two neural ODE models arising directly from the state space view and four neural network models arising from EDMD approximations of the Koopman operator (with learned dictionaries).  These model variations are described in Sec. \ref{sec:model_struct} below and summarized in Table \ref{tab:knode_models}.  
Then, in Sections \ref{sec:lorenz} and \ref{sec:MFE} we present a performance comparison on two chaotic systems: the Lorenz system and a low-dimensional model of a turbulent shear flow.  
These examples demonstrate that while the standard, linear EDMD-DL fails in both cases, EDMD-DL with projection performs equivalently to the neural ODE models, which perform quite well across several metrics. This further supports our point that adding the projection step to the EDMD-DL time evolution makes the EDMD-DL model equivalent to a neural ODE, as it results in a nonlinear, neural network representation of the dynamics on the state space, rather than a linear representation of the dynamics on the space of observable functions as is typically associated with Koopman-based approaches.

While the examples presented here are low-dimensional, previous works have shown that neural ODE models \cite{Linot2022,Perez2023,Linot2023b,ConstanteAmores2024a} and Koopman-based models with state-space projection \cite{constante-amores_enhancing_2024, ConstanteAmores2024, constante-amores_data-driven_2024} can be applied to higher dimensional systems, including chaotic PDEs such as the Kuramoto-Sivashinsky system, Kolmogorov flow, and turbulent Couette and pipe flows.  
For such high-dimensional systems, these studies paired dynamical models with autoencoder-based dimension reduction, learning the dynamics of a latent representation given by an autoencoder.  
In Ref. \onlinecite{Linot2022}, it was shown that developing neural ODE models on such reduced representations can lead to better predictive performance than training a neural ODE directly on the higher dimensional data.  
Since the purpose of the numerical examples here is to demonstrate the equivalence of EDMD-DL with a state space projection to a neural ODE, we have chosen examples with complex, chaotic dynamics, but which are still sufficiently low dimensional so that additional dimension reduction steps are not needed.  This allows us to assess the predictive capabilities of the dynamical models without introducing additional errors due to dimension reduction.

\subsection{Model structures}\label{sec:model_struct}

\paragraph*{Neural ODE Models}
 The first neural ODE model considered is a basic neural ODE, without any explicit structural constraints.  This model approximates the right hand side of the ODE in Eq. \ref{eq:ode} with a neural network, as described in Sec. \ref{sec:neural_odes} and is trained to minimize the one-step state-prediction loss in Eq. \ref{eq:node_objective}. 
We also consider an \emph{EDMD-structured neural ODE}, in which the output of the neural network is structured to represent the state space dynamics similarly to EDMD-based approaches.  In this model, the neural network maps the state $\bx$ to a high-dimensional feature vector that explicitly contains the state (as in Eq. \ref{eq:state_in_dict}) before a final linear layer that selects a linear combination of these features to represent the vector field.  This neural ODE therefore has the same structure as the model in Eq. \ref{eq:edmd_node}, but is trained using the neural ODE training procedure, rather than any form of EDMD. 

\paragraph*{Discrete-time EDMD models}
The first EDMD approach that we consider is the standard EDMD-DL method, where a dictionary parametrized by a neural network and the corresponding Koopman operator matrix representation are jointly learned by optimizing the objective in Eq. \ref{eq:edmd_loss}.  Then to model the time-evolution, the initial state is lifted and propagated linearly in the lifted space using the Koopman matrix (see Eq. \ref{eq:linprop_psi_i}). 
We also consider the EDMD variation with projection, where the same dictionary and Koopman matrix are used, but where on each timestep, the state is lifted, evolved forward in time, and then reprojected (see. \eqref{eq:nonlinprop_psi_i}).  
In each EDMD-DL implementation, we constrain the dictionary to be of the form in \eqref{eq:state_in_dict}, where the state is explicitly included as a nontrainable set of dictionary elements.  

\paragraph*{Continuous-time EDMD models} 
We consider two variations of continuous-time models derived from EDMD.  
In the first, we use the dictionary obtained from EDMD-DL as the neural network in a neural ODE.  That is, the dictionary is optimized, as in EDMD-DL, to minimize \eqref{eq:edmd_loss}, and then the linear layer that maps to the output is optimized through the neural ODE training procedure.  
So, the model has the structure of \eqref{eq:edmd_node}, where the dictionary $\Psi(\bx; \theta_\Psi)$ is determined by EDMD-DL and the linear mapping $\bA\equiv\bP\bL\trans$ is trained as with the neural ODE loss in \eqref{eq:node_objective}. 
We refer to this variation as an \emph{EDMD basis neural ODE}. 
Lastly, we consider a neural ODE model that is obtained directly from EDMD-DL, without a separate training procedure.  That is, the matrix representation of the generator $\bL$ is obtained from the EDMD-DL Koopman matrix $\bK_\tau$ by a matrix logarithm, as in \eqref{eq:gedmd_log}.  
We refer to this variation as an \emph{EDMD direct neural ODE}. 

\begin{figure*}
    \centering
    \hfill
    \begin{minipage}[c]{0.5\linewidth}
        \centering
        \includegraphics[width=\linewidth]{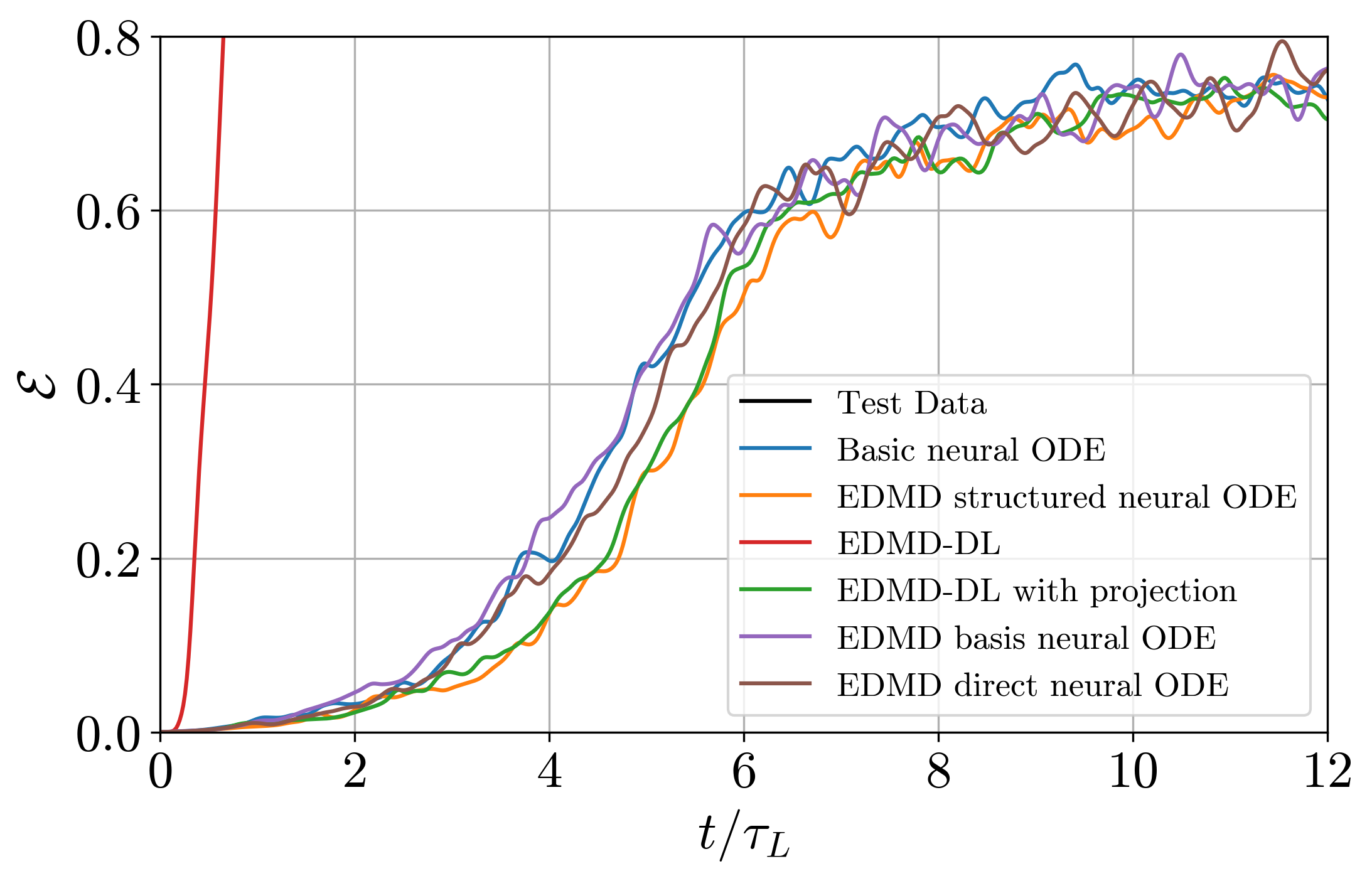}
        \\
        (a)
    \end{minipage}%
    \hfill
    \begin{minipage}[c]{0.4\linewidth}
        \centering
        \includegraphics[width=\linewidth]{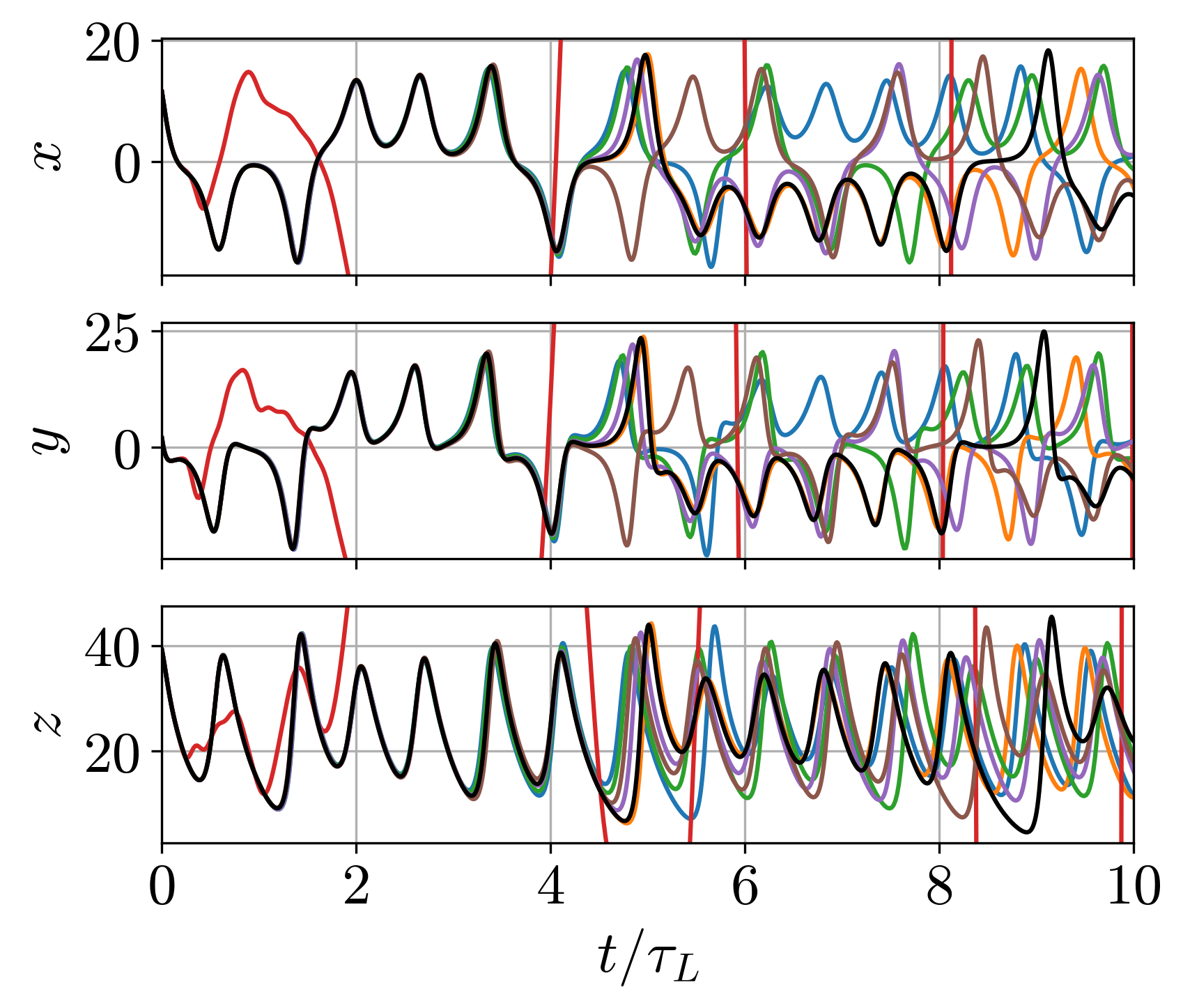}
        \\
        (b)
    \end{minipage}
    \hfill
    \caption{Prediction results for the data-driven models on the Lorenz system for sampling interval $\tau=0.02$. (a) Normalized ensemble-averaged error over time. (b) Representative trajectory timeseries of the states with predictions from each model.}
    \label{fig:lorenz}
\end{figure*}

In the sections below, we provide a numerical comparison of the performance of these methods on two example systems:  dynamics on the attractor of the Lorenz system and on a 9-mode model for a turbulent shear flow \cite{Moehlis2004,Moehlis2005}.  
In each implementation, we have made an effort to keep the neural network architectures, training procedures, and hyperparameters as consistent as possible between the different variations.  
In particular, all of the neural networks have 4 hidden layers with 200 nodes in each hidden layer and GELU activations between the layers.  
For the EDMD-DL models, the network output representing the nonlinear elements of the dictionary is of dimension 200, so that the total dictionary size is $k=200+n$, where $n$ is the state dimension. 
The models are implemented in PyTorch and trained using the AdamW optimizer with weight decay of $10^{-6}$ and a OneCycle learning rate schedule which peaks at $10^{-3}$ after $30\%$ of the total number of training iterations and then decays to $10^{-5}$.

\subsection{Lorenz system}\label{sec:lorenz}
We consider the Lorenz system  
\begin{subequations}
    \begin{align}
        \frac{dx}{dt} &= \sigma (y-x)\\
        \frac{dy}{dt} &= x(\rho-z) - y \\ 
        \frac{dz}{dt} &= xy - \beta z
    \end{align}
\end{subequations}
with parameter values $\rho=28$, $\sigma = 10$, and $\beta = 8/3$. 
For these parameter values, the Lyapunov timescale is $\tau_L \approx 1.1$ \cite{viswanath1998lyapunov}.
A training dataset is generated by integrating the system numerically using a Dormand-Prince (4)5 ODE solver and storing the solution with sampling time of $0.02$.  
The first 1000 time units are neglected as transience so that the data lies on the attractor, and the next $50,000$ time units are stored with an $80/20$ train/test split. 
All models are trained for $100,000$ training iterations with a batch size of $40$.   
We also study the effects of the training timestep, $\tau$, by training models with $\tau$ varying from $0.02$ to $0.1$.  

Figure \ref{fig:lorenz} demonstrates the short-time tracking error for each of the models on the Lorenz system, with $\tau=0.02$.  
Fig. \ref{fig:lorenz}(a) shows an ensemble-averaged relative error, 
\begin{equation}
    \calE(t) = \left\langle\frac{ \|\hat{\bx}(t) - \bx(t)\|}{\|\bx(t)\|}\right\rangle
\end{equation}
where $\hat{\bx}$ is the model prediction and $\langle\cdot\rangle$ denotes an ensemble average.  We evaluate this error over time for each of the model by simulating the models from $1,000$ randomly selected initial conditions from the test dataset.  
Fig. \ref{fig:lorenz}(b) shows an example trajectory timeseries, along with the predictions from each of the models.  
From these results, it is clear that the standard linear EDMD time evolution prediction diverges rapidly from the true trajectory, while the other models perform well, with less than $2\%$ relative error at $t=\tau_L$.  
This failure of the linear EDMD-DL model is likely due to the fact that it is not possible to find a finite-dimensional linear representation
of a system with essentially nonlinear characteristics, such as the chaotic behavior of the Lorenz attractor \cite{brunton_koopman_2016,kaiser_data-driven_2021,Page2019,cenedese_data-driven_2022,haller_data-driven_2024}. 
In contrast to the failure of linear EDMD-DL, EDMD-DL with projection performs well -- equivalently to the neural ODE models.  This equivalent performance underscores our main point that the state space projection makes the EDMD-DL model equivalent to a neural ODE in that it gives a nonlinear, neural network representation of the dynamics on the state space. 
From the ensemble-averaged error, we also see that the EDMD with projection and the EDMD-structured neural ODE have a slightly slower error growth than the basic neural ODE, indicating that there are benefits to explicitly including the state in the dictionary, which explicitly separates the linear and nonlinear portions of the neural ODE model.  

\begin{figure*}
    \centering
    \includegraphics[width=0.85\linewidth]{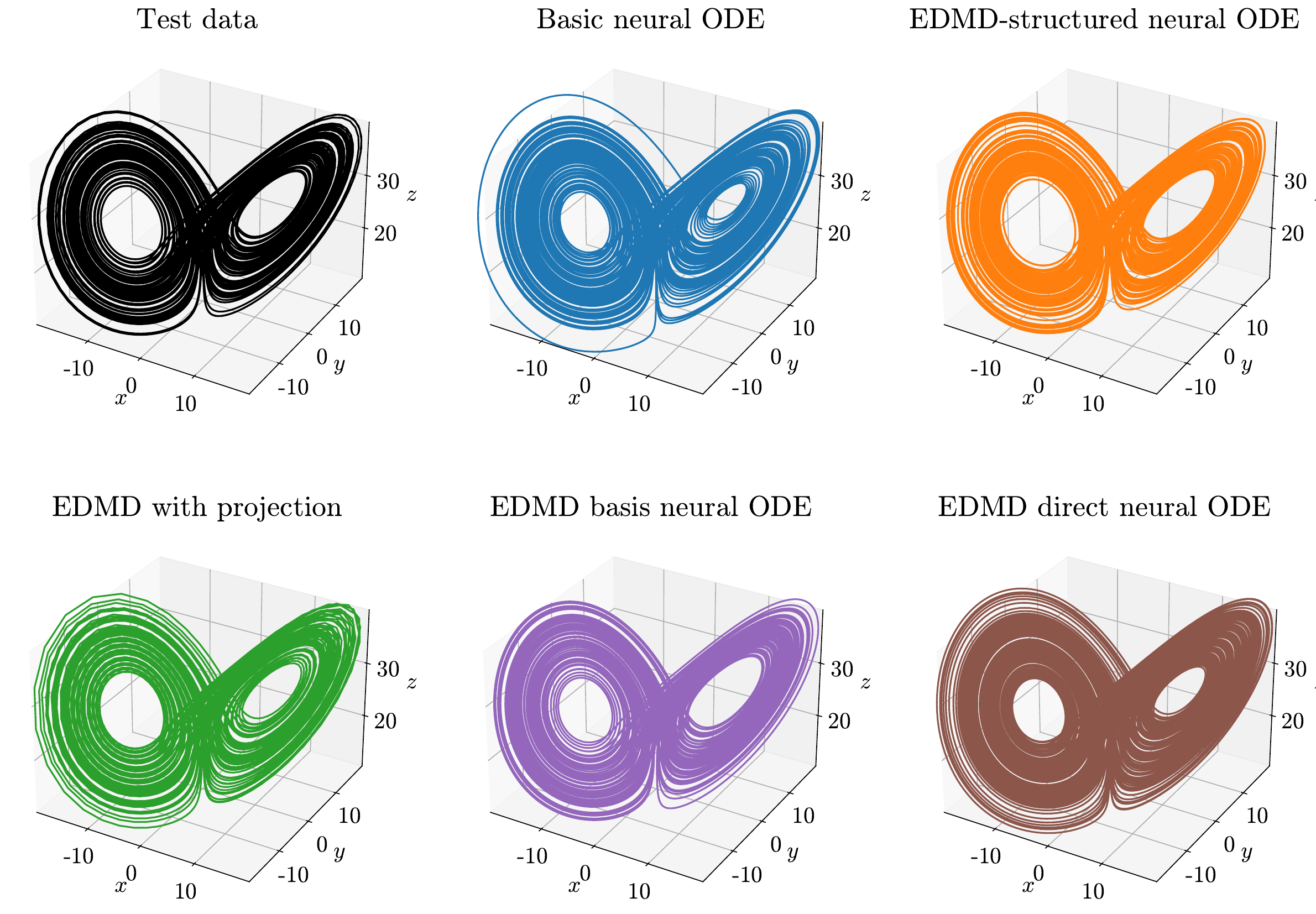}
    \caption{Long time predictions of each model from a common initial condition for $100$ time units on the Lorenz system. Each of these models maintains the shape of the attractor for long times. These trajectories are generated by models trained with timestep $\tau=0.02$.}
    \label{fig:lorenz_3d}
\end{figure*}

\begin{figure}
    \centering
    \includegraphics[width=0.9\linewidth]{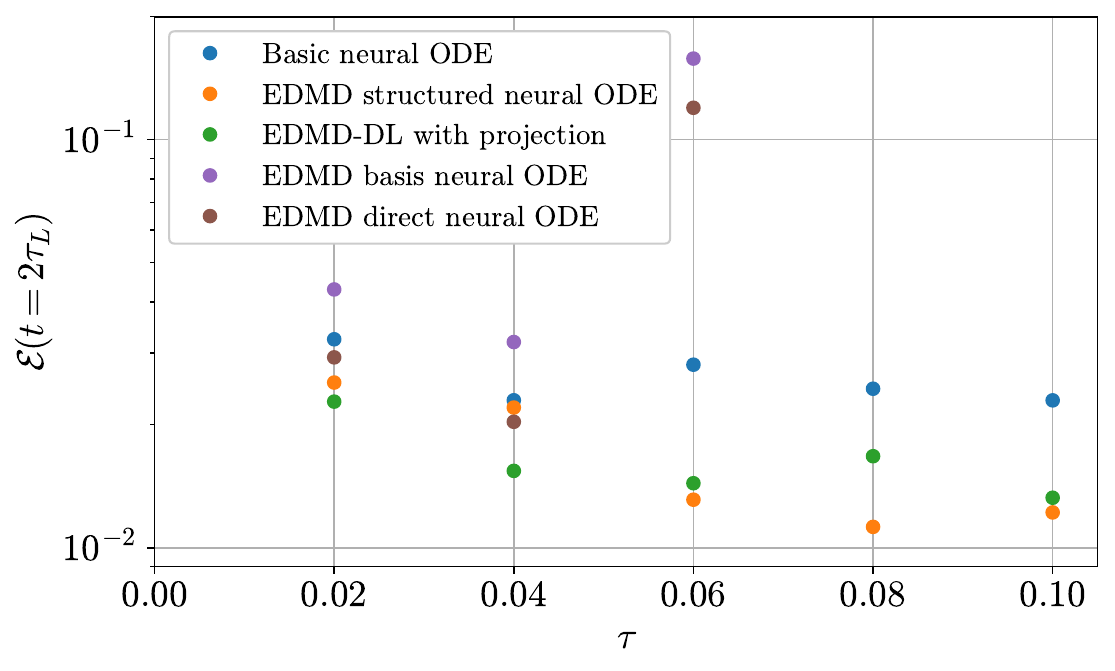}
    \caption{Normalized ensemble-averaged error of each model evaluated  at $t=2$ for varying training data sampling time, $\tau$. }
    \label{fig:lorenz_err_v_tau}
\end{figure}

We now compare the performance of the models when trained on data sampled at larger sampling times. 
For this, starting with the original training dataset we downsample it in time to effectively vary the sampling time interval, $\tau$. 
We train new models on the downsampled datasets, while holding the total number of training iterations and other training parameters constant.  We then compare the models in terms of the ensemble-averaged error, $\calE$, evaluated at $t=2\tau_L$.  
The results are shown in Fig. \ref{fig:lorenz_err_v_tau}. 
We see that at large sampling times, the EDMD basis neural ODE and EDMD direct neural ODE begin to fail, while at small sampling times (less than $\tau = 0.06$), all of the models perform well, with the relative error ranging from $1\%$ to $5\%$ at a prediction horizon of $2\tau_L$. The neural ODE models and EDMD-DL with projection maintain this level of accuracy even at larger sampling times up to $\tau=0.1$. 

Finally, while Figs. \ref{fig:lorenz} and \ref{fig:lorenz_err_v_tau} analyzed the short-time performance of these models, we demonstrate the long-time performance of the models by simulating each model for $100$ time units from a common initial condition.  The resulting trajectory from each model is shown in Fig. \ref{fig:lorenz_3d}.  We see that each of the models considered (excluding the standard EDMD-DL approach) are able to capture the dynamics and reproduce the expected butterfly shape of the Lorenz attractor.

\subsection{Low dimensional model of turbulent shear flow} \label{sec:MFE}
Next we compare these methods on the nine-mode model for a sinusoidal shear flow between parallel plates developed by Moehlis, Faisst, and Eckhardt (MFE) in \cite{Moehlis2004,Moehlis2005}.  This model provides a useful test case, as it displays complex dynamics with many characteristics associated with turbulent shear flows in the transition regime, such as long chaotic transient intervals with rare quasi-laminarization events, with all trajectories eventually collapsing to the laminar state.  
These complex dynamics allow us to assess not only a model's ability for short time predictions and reproduction of long time statistics in the turbulent region, but also the accuracy of the lifetime predictions and the model's ability to predict extreme events.  
For these reasons, this model has become a common test-case for data-driven time-series prediction methods, particularly in the fluid dynamics community \cite{Srinivasan2019, Racca2022, Fox2023, constante-amores_data-driven_2024}. 

\begin{figure}
    \centering
    \includegraphics[width=\linewidth]{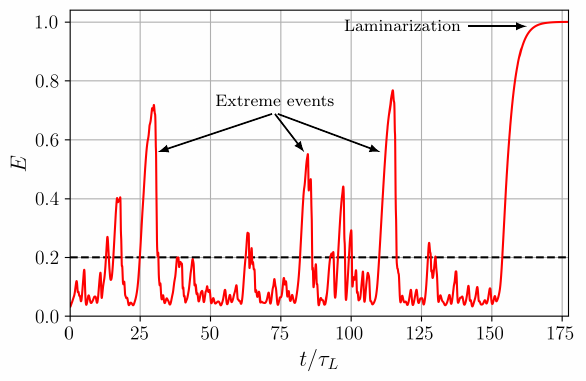}
    \caption{Typical time series for the energy in the MFE model for a turbulent shear flow. The horizontal dashed line depicts the energy threshold used to define an extreme event.   }
    \label{fig:MFE_energy}
\end{figure}

The domain is wall-bounded at $y=\pm1$ and periodic in the streamwise and spanwise directions ($x$ and $z$), with lengths $L_x=4\pi$ and $L_z=2\pi$, respectively; and we consider a channel Reynolds number of $\mathrm{Re}=400$, following Refs. \onlinecite{Srinivasan2019, Racca2022,Fox2023,constante-amores_data-driven_2024}.  
The Lyapunov timescale associated with these parameter values is $\tau_L \approx 61$ \cite{Racca2022}. 
The nine mode model is constructed as a Galerkin projection of the Navier-Stokes equations onto nine Fourier modes, as described in Ref. \onlinecite{Moehlis2004}.  
That is, the fluid velocity field $\bu(\bx, t)$ is approximated as a superposition of the modes, $\bu_i(\bx)$, with time-varying amplitudes $a_i(t)$
\begin{equation}
    \bu(\bx, t) = \sum_{i=1}^9 a_i(t) \bu_i(\bx)\,. 
\end{equation}
Performing the Galerkin projection yields a set of nine ODEs for the amplitudes.  These ODEs, as well as the Fourier modes for the model, are given explicitly in Ref. \onlinecite{Moehlis2004}. 
The first mode of the model, $\bu_1(\bx)$, represents the base flow profile, so the laminar flow state is associated with the amplitudes $\ba = [1,0,0,0,0,0,0,0,0]\trans$. 
The flow can be characterized using the energy $E(t) = \sum_{i=1}^9a_i^2(t)$. 
In the turbulent region of the state space, the system exhibits extreme, high energy quasi-laminarization events, while the laminar state corresponds to the steady value of $E=1$. 
A typical time-series of the energy for this system is shown in Fig. \ref{fig:MFE_energy}

\begin{figure*}
    \centering
    \begin{minipage}[c]{0.5\linewidth}
        \centering
        \includegraphics[width=\linewidth]{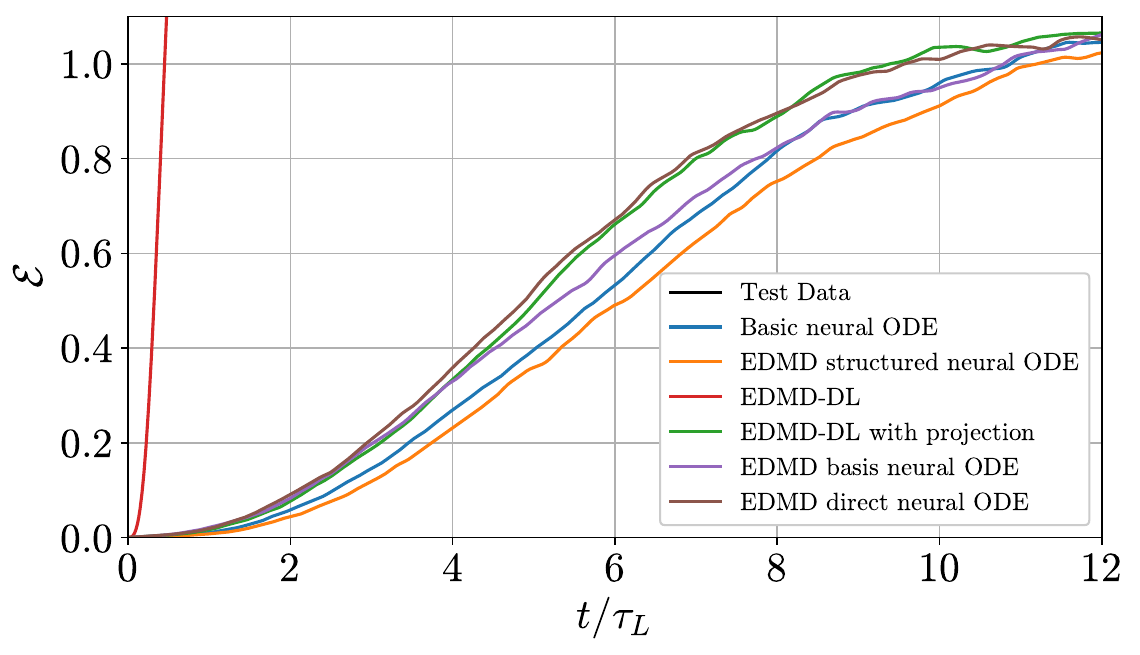}
        \\
        (a)
    \end{minipage}%
    \hfill
    \begin{minipage}[c]{0.4\linewidth}
        \centering
        \includegraphics[width=\linewidth]{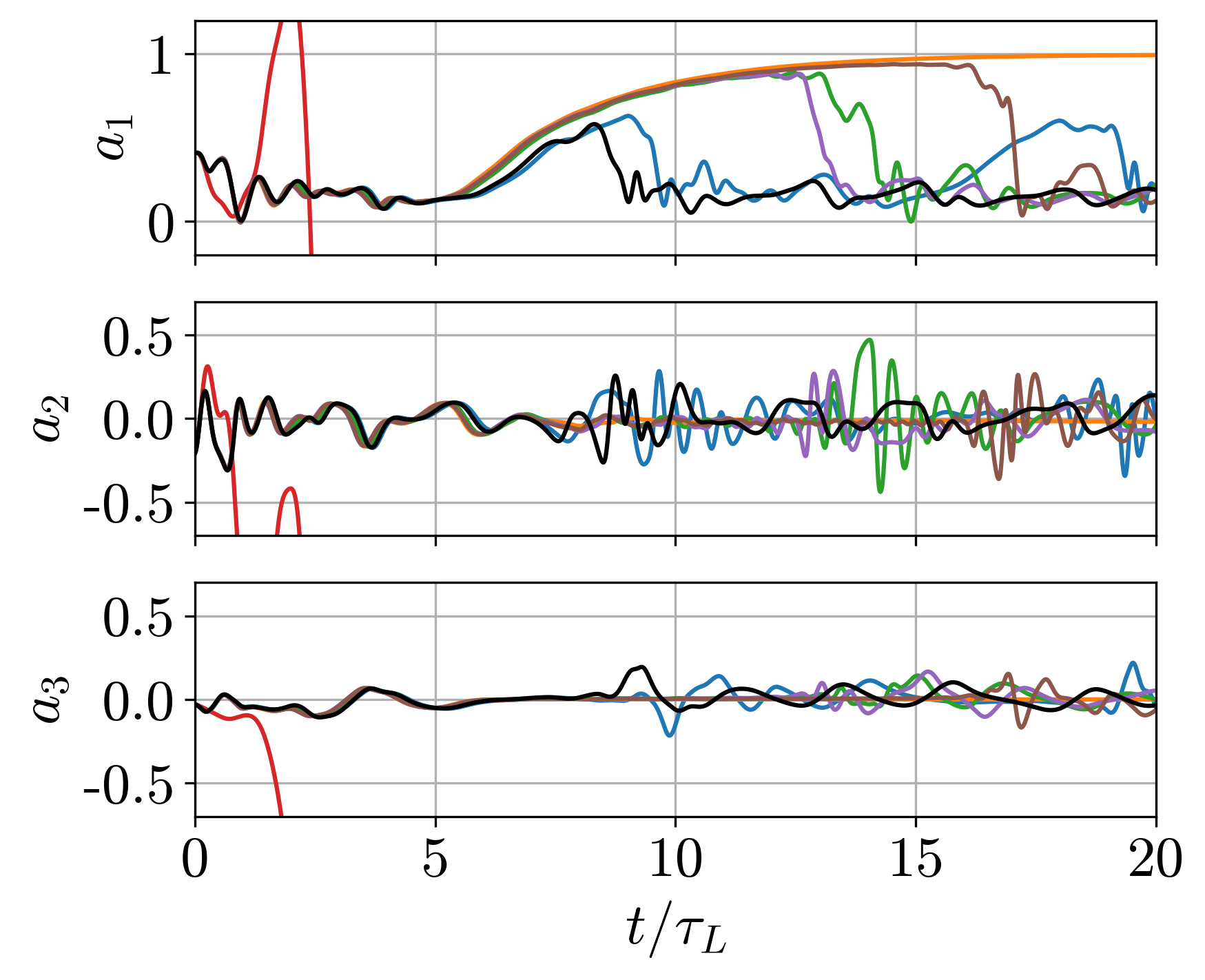}
        \\
        (b)
    \end{minipage}
    \caption{Prediction results for the data-driven models on the MFE model of a turbulent shear flow.  (a) Normalized ensemble-averaged error.  (b) Representative trajectory timeseries of the first three modes.  }
    \label{fig:MFE_tracking}
\end{figure*}

A training dataset is constructed by numerically integrating the system from 100 random initial conditions, $\ba(0)\sim\mathcal{U}([-0.1,0.1]^9)$ for 30,000 time units or until laminarization occurs with data stored at time intervals of $\tau=0.5$. 
The first $500$ time units of each trajectory are discarded as transience.  
These trajectories have a mean lifetime of approximately $1.1\times10^4$ (or $180\,\tau_L$) and the resulting dataset contains approximately $2.1 \times 10^6$ snapshots.  
A test dataset is generated using an identical procedure, using another set of randomly selected initial conditions.  
All models are trained for 10 epochs with batch size of 40 ($5.25\times 10^5$ iterations).

Figure \ref{fig:MFE_tracking} shows the short-time prediction results for the data-driven models on the MFE system.  Fig. \ref{fig:MFE_tracking}(a) shows the ensemble-averaged error over time for each model, computed (as before) by simulating the models from 1000 randomly selected initial conditions from the test dataset.  
Fig. \ref{fig:MFE_tracking}(b) shows a representative trajectory from the test dataset, along with the model predictions for the first three mode amplitudes.  
As in the case of the Lorenz system, we see that the standard EDMD time evolution diverges rapidly from the true trajectory, while the other time evolution methods perform well.
Here all of the models trained with the neural ODE training procedure 
 outperform all of the EDMD-DL-based models in terms of ensemble-averaged relative error, with the EDMD-structured neural ODE performing slightly better than the basic neural ODE.  
Nevertheless, that all of these models perform quite well, with less than $3\%$ relative error at $t=\tau_L$.  

\begin{figure*}
    \centering
    \includegraphics[width=0.9\linewidth]{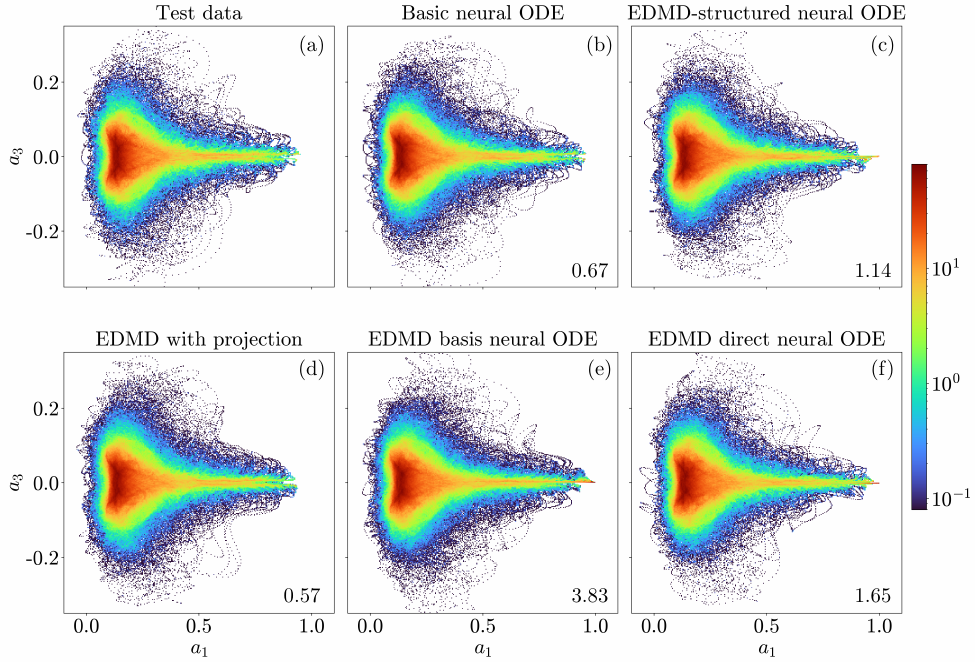}
    \caption{Joint probability densities of amplitudes $a_1$ and $a_3$ for (a) the test data set, (b) the basic neural ODE model, (c) the EDMD-structured neural ODE model, (d) EDMD with projection (e) the EDMD basis neural ODE, and (f) the EDMD direct neural ODE. The inset values give normalized Wasserstein distance  $\overline{\calW}_2$ between the distribution predicted by the model and the test data distribution. Note the logarithmic scale.  }
    \label{fig:MFE_longstats}
\end{figure*}

To assess the accuracy of the models in terms of reproducing the long-time statistics of the system, we simulate the models from the 100 initial conditions of the trajectories in the test dataset for 30,000 timesteps or until relaminarization is predicted.  
Shown in Fig. \ref{fig:MFE_longstats} are the joint probability density functions (PDFs) of the amplitudes $a_1$ and $a_3$, from this test data set and as predicted by each of the models.  The standard EDMD model is omitted here, since it tends to rapidly diverge from the true trajectory, as shown in Fig. \ref{fig:MFE_tracking}. From Fig. \ref{fig:MFE_longstats}, it is apparent that each of the other models reproduces the statistics qualitatively well.  
To quantify the accuracy of these long-time statistics, we compute a Wasserstein distance (a.k.a. earth mover's distance) to quantify the discrepancy between the predicted and true joint PDFs.  
For this, we consider the empirical distribution of the test distribution as the true joint PDF. 
The Wasserstein distance measures the distance between distributions by solving an optimal tranport problem to determine the most efficient way to move mass between distributions.  
Given two discrete distributions, $\mu$ and $\nu$ (as obtained from a normalized histogram of the data using $n_\mu$ and $n_\nu$ bins, respectively), the Wasserstein-2 distance, $\calW_2(\mu,\nu)$ between the distributions is given by 
\begin{equation}
\begin{split}
    \mathcal{W}_2(\mu, \nu) = \left(
    \min_{\gamma} \sum_{i,j} \gamma_{i,j} \|\mu^{(i)} - \nu^{(j)}\|_2^2
    \right)^{\frac{1}{2}} \\
    \text{s.t. } \gamma \mathbf{1} = \mu, \quad \gamma\trans \mathbf{1} = \nu, \quad \gamma \geq 0
\end{split}
\end{equation}
where $\gamma\in\reals^{n_\mu\times n_\nu}$ is a transport plan (i.e., $\gamma_{i,j}$ assigns the amount of mass to move from bin $\mu^{(i)}$ to bin $\nu^{(j)}$).  
Other distances for quanitatively comparing distributions exist, such as the Kullback-Leibler divergence or a norm of the error, but these distances are typically more sensitive to small shifts of the distribution or the discretization of the distribution than the Wasserstein distance. 
For a more detailed discussion of the use of the Wasserstein distance and a comparison of its properties to those of other statistical divergences commonly used in data science, we refer to Peyr\'e and Cuturi \cite{peyre2019computational}. 
We compute the Wasserstein distances using the python library, Python Optimal Transport \cite{flamary2021pot}, and we report the distances normalized by the distance between the distributions of the test and training datasets.  That is, we report the normalized Wasserstein distance
\begin{equation}
    \overline{\calW}_2(\mupred,\mutest) = \frac{\calW_2(\mupred, \mutest)}{\calW_2(\mutrain,\mutest)}. 
\end{equation}
for each model where $\mupred$ is the distribution of the dataset generated by a given model, and similarly $\mu_\mathrm{ test}$ and $\mu_\mathrm{train}$ are the distributions of the test and training datasets respectively.  
So $\overline{\calW}_2=1$ indicates that the error between the predicted and test histograms is of the same magnitude as the error between the test and training datasets, which are generated by the same model with different initial conditions.  
The $\overline{\calW}_2$ values for each model are displayed as an inset in the joint PDF plots in Fig. \ref{fig:MFE_longstats}. 
We see that each model has a $\overline{\calW}_2$ value of approximately 1 or less, with the exception of the EDMD basis neural ODE model, which appears to suffer from a large error near to the laminar state, indicating poor prediction of relaminarization, while otherwise appearing qualitatively accurate.

\begin{figure}[t]
    \centering
    \includegraphics[width=\linewidth]{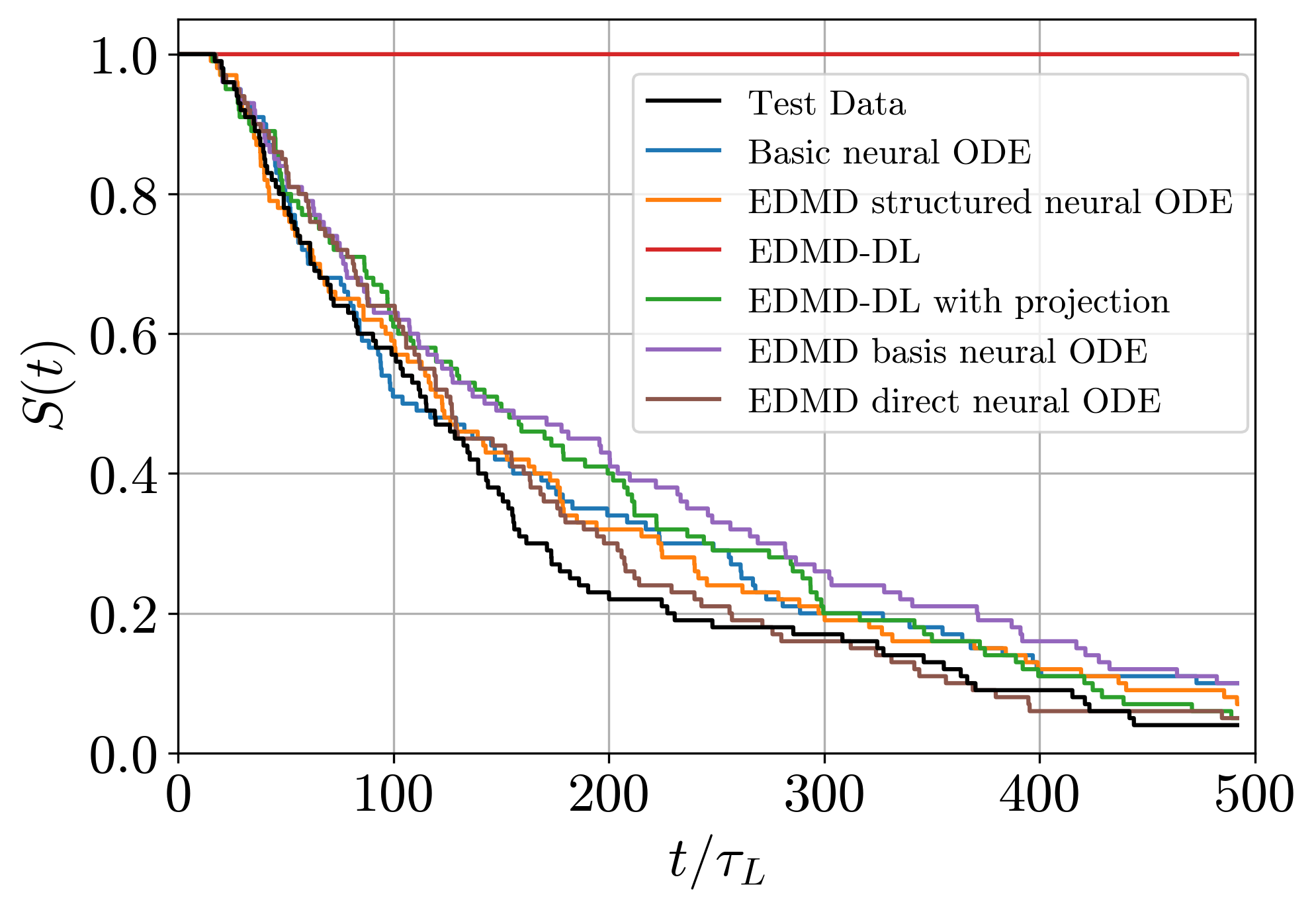}
    \caption{Lifetime distribution predictions for the models, given in terms of the survival function $S(t)$. }
    \label{fig:MFE_lifetime}
\end{figure}

Using the same predicted and test datasets as used for quantifying the long-time statistics, we also evaluate the models in terms of their ability to predict the lifetime distribution for the system.  
The lifetime distribution describes the expected amount of time that trajectories spend in the turbulent region of the state space before laminarization.  
The lifetime distribution is quantified by the survival probability, $S(t)$, which describes the probability that a trajectory will still be turbulent after time $t$ \cite{Moehlis2004}. 
For this, we say that the system has laminarized when it reaches a high-energy, near-laminar state ($E=\sum_{i=1}^9a_i^2 > 0.8$) which remains steady for 5 time units (to within a tolerance of $10^{-6}$).  
Fig. \ref{fig:MFE_lifetime} shows the lifetime distribution of the test dataset along with the lifetime distribution predicted by each of the models.  
From the lifetime distribution of the test-dataset, it can be seen that trajectories tend to remain in the turbulent region for very long times, as the half-life of the survival function is over 100 Lyapunov times. Therefore, to produce accurate reconstructions of the long-time statistics, as in Fig. \ref{fig:MFE_longstats}, the models are required to maintain accurate predictions over very long timescales.  
In Fig. \ref{fig:MFE_lifetime}, we see that each of the considered models, aside from the standard EDMD perform comparably, with each of them accurately approximating the lifetime distribution.
The basic neural ODE and EDMD-structured neural ODE models appear to capture the distribution the best at shorter times ($t < 100 \tau_L$), while the EDMD-direct neural ODE performs better at larger times where the other models tend to overestimate the survival probability.  
The standard EDMD-DL model prediction tends to grow rapidly, and never collapses to the laminar state, which results in a survival probability of $1$ by the definition above.  

Finally, we assess the ability of the data-driven models to forecast rare quasi-laminarization events within a short time horizon by casting this task as a binary classification problem, following Racca and Magri \cite{Racca2022}.  
That is, we evaluate whether extreme events predicted by the model within a specified time interval are true positives ($TP$) or false positives ($FP$).  Similarly, when a model does not predict an extreme event in the interval, we evaluate whether this is a true negative ($TN$) or a false negative ($FN$).  
In this way, we can also evaluate how rapidly the quality of extreme event prediction degrades as the prediction horizon, $h$, before the interval of interest is increased.  
There are multiple ways to quantify the statistical accuracy of predictions in binary classification problems. 
Precision quantifies the fraction of true positive predictions out of all positive predictions, $P=TP/(TP+FP)$, while recall quantifies the ratio of true positive predictions to the total number of event occurences, $R=TP/(TP+FN)$.  
Here we report the $F$-score, which is given by the harmonic mean of precision and recall
\begin{equation}
    F = \frac{2}{P^{-1}+R^{-1}} = \left(1+\frac{FP+FN}{2TP}\right)^{-1}.
\end{equation}
To compute the $F$-score for our models, we first select a set of starting points from the test dataset evenly spaced at 1 $\tau_L$ apart.  From these starting points, we simulate the model forward for a time of $h+\tau_L$, and assess the model's extreme event prediction within the interval $[h, h+\tau_L]$. 
By processing the test dataset in this way, we construct a dataset containing 18,000 starting points (and intervals) with 947 quasi-laminarization events.  
We say that a quasi-laminarization event has occurred within a given interval if the energy is below the threshold value at the beginning of the interval and increases past the threshold within the interval. 
Some examples of these quasi-laminarization events are depicted in Fig. \ref{fig:MFE_energy}, along with the energy threshold considered here.  
We use an energy threshold of $E=0.2$ and consider prediction horizons ranging from $0.5\tau_L$ to $4.5\tau_L$. 

\begin{figure}[t]
    \centering
    \includegraphics[width=\linewidth]{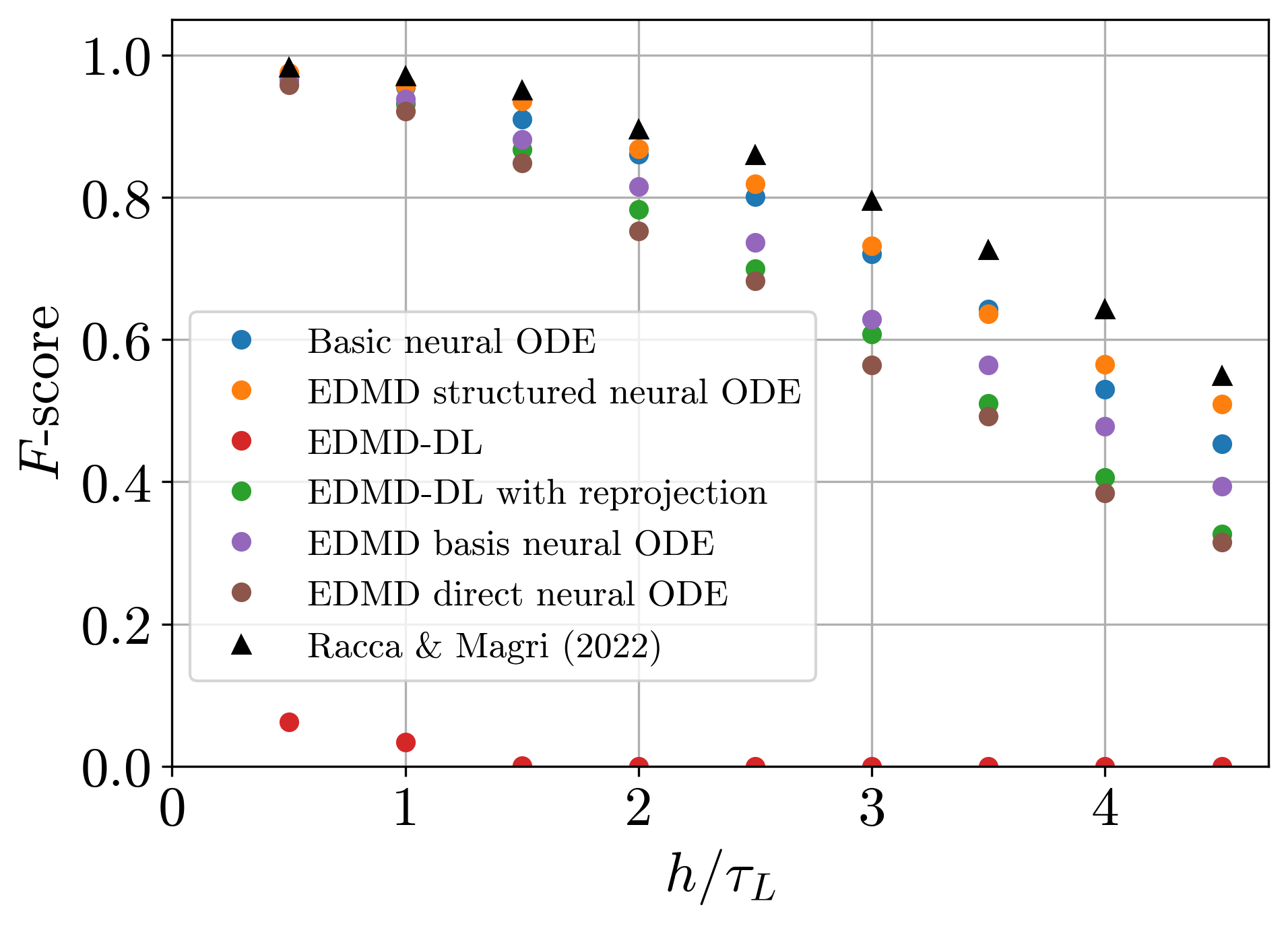}
    \caption{$F$-score assessment of quasi-laminarization event prediction as a function of prediction horizon, $h$. }
    \label{fig:MFE_fscore}
\end{figure}

Figure \ref{fig:MFE_fscore} shows the $F$-score as a function of the prediction horizon normalized with Lyapunov time, $h/\tau_L$, for each of the models.  From these results, it is clear that the continuous time neural ODE models outperform the discrete time and EDMD-DL-based approaches, more accurately forecasting extreme events for every prediction horizon.   
Among the EDMD-DL-based models, the EDMD-basis neural ODE outperforms EDMD-DL with projection and the EDMD direct neural ODE. 
The standard EDMD-DL prediction is poor for every prediction horizon, resulting in a much lower $F$-score than the other models. 
Also included for comparison in Fig. \ref{fig:MFE_fscore} are the $F$-score results from Racca and Magri \cite{Racca2022}, who developed a model using an echo state network, a type of recurrent neural network. Its predictions are based on long sequences of past states and are thus non-Markovian, while our model predictions are based only on the initial state.   Nevertheless, their model only slightly outperforms the present models, especially for the case of the EDMD-structured neural ODE. 

While these $F$-score results results demonstrate the short-time prediction of extreme events, next we consider the long-time statistical accuracy of extreme-event prediction for each of these models.  For this, we compute
the ergodic average of an extreme event indicator function.  
This indicator function is defined as 
\begin{equation}
          I_{EE}(\mathbf{a}) = \begin{cases}
            1 \qquad E(\mathbf{a}) > 0.2\\
            0 \qquad E(\mathbf{a})  \leq 0.2
        \end{cases}
\end{equation}
where $E(\mathbf{a})$ is the kinetic energy, as defined above. 
The ergodic average of this indicator essentially represents the fraction of time that trajectories spend in the high kinetic energy, near-laminar region of the state space.  
The ergodic average of the extreme event indicator function for each of the models is computed over the trajectory datasets used in Fig. \ref{fig:MFE_longstats}, which consist of simulations from each model from 100 initial conditions for 30,000 timesteps or until relaminarization is predicted.
The resulting ergodic averages are given in Table \ref{tab:indicator_mean}.
     \begin{table}
         \centering
         \caption{Ergodic average of extreme-event indicator function.}
         \begin{tabular}{|l|c|} 
         \hline Model & $\langle I_{EE} \rangle$\\
         \hline \hline 
         Test data & ~0.130\\
         Basic neural ODE & 0.125\\ 
         EDMD structured neural ODE ~~& 0.131\\
         EDMD-DL with projection & 0.128\\
         EDMD basis neural ODE & 0.142\\
         EDMD direct neural ODE & 0.131\\
         \hline \end{tabular}
         \label{tab:indicator_mean}
     \end{table}
    This result further demonstrates that EDMD-DL with projection and each of the neural ODE models perform comparably, yielding an ergodic average of this indicator that is close to the value when calculated on the test data. 
     This indicates that these models successfully capture the long-time statistics of extreme events. 

\section{Conclusions}
Developing and improving upon neural network-based methods for time evolution prediction of dynamical systems is a promising approach for the forecasting of complex dynamical systems.  
In this work, we have examined the parallels between neural ODE methods and methods based on finite-dimensional approximations of the Koopman operator using neural network-based dictionaries.  
While methods based on the Koopman operator typically lead to an approximation of linear dynamics on the space of observables, in practice these linear methods often provide poor predictions of the time-evolution of the states due to the difficulty associated with determining a dictionary which spans a Koopman invariant space for a given system.  
This is especially true for chaotic systems, such as the ones considered here. 
It has been seen that the state-predictions of these Koopman-based methods can be greatly improved by projecting back to the state space on each timestep.  
Here we have pointed out that by relifting to the observable space from the state space on each step, a nonlinearity is introduced into the EDMD time evolution formulation, resulting in a nonlinear discrete-time map on the state space.  
Therefore using an EDMD approximation of the operator in this way with a neural network dictionary can be seen as a discrete-time version of a neural ODE.  
Furthermore, we can also apply a similar projection strategy to the continuous time formulations of EDMD based on the Koopman generator to obtain a continuous time neural ODE directly from the EDMD-DL-optimized operator and dictionary without any further training.  
Additionally, the model structures arising from EDMD-DL with projection naturally possess a neural network structure that is commonly used in neural ODEs, in which the state is first expanded out to a high-dimensional feature vector (represented by a multi-layer neural network) followed by a linear mapping back to the state space to yield the vector field of the dynamics. 

While including the state space projection step in the EDMD-DL timestepping procedure does reduce the model to a neural network representation of the dynamics on the state space, the training procedure and structure of the model are the same as in standard EDMD-DL.  
Therefore, the model inherits the benefits of the EDMD-DL method.  
It provides a matrix approximation of the Koopman operator, which can be used to study Koopman eigenvalues, eigenfunctions, and modes \cite{williams_datadriven_2015,li_extended_2017}. 
This has been demonstrated in the original EDMD-DL paper \cite{li_extended_2017}, where it was also shown that optimizing the basis in EDMD-DL typically yields more accurate and efficient representations of the operator and its spectrum than alternative choices of basis, such as radial basis functions or Hermite polynomials.   
Many previous works have shown that analysis of Koopman eigenfunctions can be used to glean physical insights into the structure underlying a dynamical system by identifying large scale structures and their associated frequencies in high-dimensional systems, parameterizing basins of attraction, and identifying invariant sets and coherent sets (see, for example \onlinecite{budisic_applied_2012,mezic_analysis_2013,rowley_model_2017,brunton_modern_2022} for detailed reviews of these applications).

In this work, we have implemented several variations of neural ODEs and EDMD-DL, developed by combining different aspects of their respective model structures and training procedures.  
We provide performance comparisons on two chaotic systems to demonstrate the equivalence of these methods and highlight the aspects of each model that lead to improved performance.   
The essentially equivalent performance between EDMD-DL with projection and the neural ODE models, in contrast to the failure of linear EDMD-DL, further emphasizes our main point: the state space projection makes the EDMD-DL model equivalent to a neural ODE in that it gives a nonlinear, neural network representation of the dynamics on the state space. 
We also show that the novel EDMD-structured neural ODE architecture
yields a small improvement in the short-time prediction performance of neural ODEs, as compared to the standard neural ODE architecture.  
In the context of neural ODEs, 
separating the linear and nonlinear parts of the vector field provides a nice structure, as the neural network only has to account for the nonlinear dynamics, with the linear terms represented separately.  In the examples considered here, the EDMD-structured neural ODE architecture, trained with the usual neural ODE loss, performed best overall, suggesting that this structure can be beneficial to the performance of neural ODEs on chaotic systems such as these.

We see that each of the neural ODE methods and EDMD-DL methods with projection tend to perform comparably on these systems in terms of short-time trajectory prediction, reconstruction of long-time statistics, and the prediction of extreme events; with all of them offering a significant improvement over the standard EDMD-DL linear time evolution on the space of observables. 
Furthermore, each of these models performs very well in terms of short-time prediction, with less than $3\%$ relative error at $t=\tau_L$ for each of the chaotic systems considered here and rare event prediction capability comparable to the non-Markovian approach of Racca and Magri \cite{Racca2022}.  
In addition to accurate short-time tracking, we have also demonstrated that the model predictions are statistically accurate for long times as well, as they yield accurate reconstructions of the long-time probability density functions on the state space and accurately predict the turbulent lifetime for the MFE model of a turbulent shear flow.  
These findings highlight the promise of neural ODEs and Koopman-based approaches for developing data-driven models capable of generating accurate time evolution predictions in complex and chaotic systems.

\section*{Acknowledgments}
This work was supported by the Office of Naval Research (grant no. N00014-18-1-2865 (Vannevar Bush Faculty Fellowship)).

\bibliography{KoopNodes}

\end{document}